\documentclass[fleqn,usenatbib]{mnras}

\usepackage[T1]{fontenc}

\DeclareRobustCommand{\VAN}[3]{#2}
\let\VANthebibliography\thebibliography
\def\thebibliography{\DeclareRobustCommand{\VAN}[3]{##3}\VANthebibliography}

\usepackage{graphicx}	
\usepackage{amsmath}	
\usepackage{amssymb}	
\usepackage{comment}
\usepackage{booktabs}

\newcommand{\Msun}{\,M$_{\odot}$}

\title[NIHAO UDGs Metalliticy]{Metallicity profiles of Ultra Diffuse Galaxies in NIHAO simulations}

\author[Cardona-Barrero et al.]{
S. Cardona-Barrero$^{1,2}$\thanks{E-mail: \href{mailto:}{salvador.cardona@iac.es}},
A. Di Cintio$^{2, 1}\thanks{Junior Leader Caixa fellow, \href{mailto:} {adicintio@iac.es}}$,
G. Battaglia$^{1,2}$, A.V. Macciò$^{3,4,5}$ $\&$ S. Taibi$^{6}$
\\
$^{1}$ Instituto de Astrof\'isica de Canarias, Calle V\'ia L\'actea s/n, E-38206 La Laguna, Tenerife, Spain \\
$^{2}$ Universidad de La Laguna Avda. Astrof\'isico Fco. S\'anchez, E-38205 La Laguna, Tenerife, Spain \\
$^{3}$ New York University Abu Dhabi, PO Box 129188 Abu Dhabi, United Arab Emirates \\
$^{4}$ Center for Astro, Particle and Planetary Physics (CAP$^3$), New York University Abu Dhabi \\
$^{5}$ Max Planck Institute für Astronomie, Königstuhl 17, D-69117 Heidelberg, Germany \\
$^{6}$ Leibniz-Institut für Astrophysik Potsdam (AIP), An der Sternwarte 16, D-14482 Potsdam, Germany
}

\date{Accepted XXX. Received YYY; in original form ZZZ}

\pubyear{2022}

\usepackage{newtxtext,newtxmath}

\begin{document}
\label{firstpage}
\pagerange{\pageref{firstpage}--\pageref{lastpage}}
\maketitle

\begin{abstract}
Supernovae feedback driven expansion has proven to be a viable mechanism to explain the average properties of Ultra Diffuse Galaxies (UDGs) such as the sizes, colors, mass and internal kinematics. 
Here, we  explore the origin of stellar metallicity gradients in feedback driven simulated UDGs from the NIHAO project and compare them with the observed distribution of metallicity gradients of both  Local Group dwarfs as well as of the recently observed UDG DF$44$.
Simulated UDGs display a large variety of metallicity profiles, showing  flat to negative gradients, similarly to what is observed in LG dwarfs, while DF$44$ data suggest a flat to positive gradient.
The variety of metallicity gradients in simulations  is set by the interplay between the radius at which star formation  occurs and the subsequent supernovae feedback driven stellar  redistribution: rotation supported systems tend to have flat metallicity profiles while dispersion supported galaxies show negative and steep profiles.

Our results suggest that UDGs are not peculiar in what regards their metallicity gradients, when compared to regular dwarfs. 
Desirably, a larger observational sample of UDGs' gradients shall be available in the future, in order to test our predictions.

\end{abstract}

\begin{keywords}
keyword1 -- keyword2 -- keyword3
\end{keywords}


\section{Introduction}

Low surface brightness (LSBs) galaxies have caught the interest of astronomers since decades \citep[e.g.][]{deBlok1996,Sandage1984,Bothun1985,Bothun1987}. However, due to their faintness, a systematic analysis of their properties and their relevance in the cosmological context has been possible only in  recent years.
Investing in new image processing techniques and better observing facilities \citep[][]{Abraham2014,Merritt2014,DarkEnergy2016,Fliri2016,Prole2018,Iodice2020,Zaritsky2021} has allowed to push further the surface brightness limits leading to the discovery of thousands of such objects \citep[see for example the recent works of ][]{Prole2019,Roman2019,Marleau2021}.

Of particular interest are the sub-sample known as Ultra Diffuse Galaxies (UDGs). Those systems first named by \cite{vanDokkum2015} (but discovered since the late $80$'s, e.g. \citealt{Impey1988}) are characterized by extremely low surface-brightness together with low light concentration due to their large half-light radii  \citep[see, ][]{Chamba2020}. 

UDGs were originally thought to form in Milky-Way mass like halos  in which  early gas removal had completely quenched them \citep{vanDokkum2015};  this hypothesis is supported by some observations of UDGs with abnormally large dynamical masses \citep[e.g. ][]{Beasley2016march,Forbes2021}. However further observations indicated that the vast majority of those galaxies have smaller masses, with dark-matter halos typical of  dwarfs \citep{Beasley2016,Sengupta2019,Gannon2020,Iodice2020}, and  with even some extreme examples  showing a deficit of dark-matter \citep[e.g. ][]{Mancera2021}.

UDGs have been mainly found in high-density environments \citep[][]{Mihos2015,Koda2015,Beasley2016march,Mancera2018,Lim2020,Lee2020} but with an increasing number of them being found in groups or in the field \citep[][]{MartinezDelgado2016,Bellazzini2017,Leisman2017,Roman2019, Barbosa2020}. It is unclear whether these difference is related with the formation mechanism of those systems or is an observational bias due to the particular definition of UDGs as it requires the measurement of physical distances \citep{Wright2021, Roman2017}. Unsurprisingly, UDGs found in clusters and UDGs found in less dense environments seem to show well differentiated properties \citep[e.g. ][]{Roman2017,Roman2017Jun,Kadowaki2021}, being the UDGs founded in clusters redder than field UDGs. 

In general, models that try to explain in-cluster UDGs, suggest that interactions with the environment can explain the extreme properties of these systems. Those can be achieved either by quenching and strangulation due to ram pressure striping. \citep{Yozin2015,Tremmel2020,Benavides2021} or by expansion due to dynamical heating \citep[][]{Carleton2019,Amorisco2019,Sales2020,Jones2021}. Either way, those mechanisms are unlikely to explain the field population of UDGs.
In this context, the models that invoke internal processes are introduced.  
\cite{Amorisco2016} showed that dwarfs with higher angular momentum are more likely to have a more expanded stellar component \citep[see also ][]{Rong2017,Liao2019}. On the other hand \cite{DiCintio2017} found that gas outflows driven by supernovae (SNae) feedback are able to expand the stellar component of the galaxy, alongside  dark matter \citep{dicintio14}, creating such low surface brightness systems \citep[see also ][]{Chan2018, Freundlich2020,brook21}. Finally, it seems that UDGs can have different merger histories than non-UDGs \citep{Wright2021,dicintio19}, being the UDGs more likely to form in halos that do not experience late mergers. 
None of these mechanisms are mutually exclusive and the combination of several of them can be the responsible of the full population of UDGs. For example, \cite{Martin2019} found that rapid star formation and galaxy-galaxy mergers were responsible for the formation of the full sample of UDGs in the Horizon-AGN simulations, while \cite{CardonaBarrero2020}  found that rotation supported UDGs from NIHAO simulations \citep{Wang2015} tend to have larger half-mass radii than dispersion supported ones, even thought stellar feedback was needed to form all of them \citep{DiCintio2017}.
 
There has been a large effort in the community to explore the different properties that emerge from the different formation mechanisms \citep[e.g. ][]{Sales2020,Carleton2021,Saifollahi2022,Trujillo-Gomez2022}. 
While several of the formation mechanisms shown in the literature
are able to explain, with more or less success, the average properties of  UDGs, the radial variation of such properties has yet to  be  explored in detail. The only relevant exception, in this regard, is  the work of \citet{brook21}, in which the authors clearly show that the slowly rising rotation curve of  gas rich UDG AGC$~ 242019$ favors formation scenarios in which internal processes, such as supernova-driven gas outflows, are acting to modify UDG profiles. 

Aside from the radial change of the circular velocity, another interesting property that could be used to discriminate amongst  formation scenarios is the variation of the stellar metallicity with radius.
The recent measurement of the first stellar metallicity gradient in an observed UDG, i.e. UDG DF$44$ \citep{Villaume2022},  has opened a new opportunity to test the different formation mechanisms. Metallicity gradients can put strong constrains in the evolution history of individual galaxies. Simulations have shown that steep gradients seem to naturally emerge in systems in which the available star forming gas is confined in the central regions of the galaxy, being the newly formed stars systematically more spatially concentrated and metal rich than the previous generation \citep{Schroyen2013,Revaz2018}. This evolution is known as "outside-in" evolution. On the other hand, dissipationless major mergers seem to efficiently mix the galaxy stellar populations flattening any pre-existing gradients \citep{Kobayashi2004,DiMatteo2009}. However, if there is residual star formation, a steep metallicity gradient can still be developed, as the merger has pushed the pre-existing stars to the outskirts effectively segregating different populations \citep{BenitezLlambay2016, Cardona-Barrero2021}.  Regarding minor accretion events, in massive galaxies dry-minor mergers are likely to create steep metallicity gradients by depositing metal poor stars in the external parts of the galaxy \citep[e.g. ][]{Cook2016}. In the dwarfs regime, is not well understood the importance of these process, as, even though minor mergers are expected to happen in a $\Lambda$CDM context, the halo occupation fraction at these low masses is not well constrained \citep[see][and references there in]{Deason2022}.

Finally, \cite{ElBadry2016} proposed that age gradients can also be developed by stellar migration due to feedback \citep[see also][]{Graus2019}. However this gradients can be quickly flattened due to late and extended star formation (SF) \citep[]{Mercado2021}.

In these contribution we aim to explore the metallicity gradients, in terms of  iron abundance, of the feedback driven Ultra-Diffuse Galaxies from the NIHAO simulation suite. The manuscript is structured as follows: in Section~\ref{sec:sims} we will give a brief description of the NIHAO simulations together with the sample selection; in Section~\ref{sec:results} we will explore the origin of the metallicity gradients in NIHAO UDGs and we will compare them with the observed gradients in LG dwarfs as well as in UDG DF44; finally in Section~\ref{sec:conclusions} we will provide a summary of our main findings.

\section{Simulations}
\label{sec:sims}

In this section we  briefly describe NIHAO simulations \citep[we refer the reader to ][for a more complete discussion]{Wang2015}.
The NIHAO project encompasses a sample of $\sim 100$ hydrodinamical cosmological zoom-in simulations that have been run using the N-body SPH solver \textsc{gasoline2} \citep{Wadsley2017} in a flat $\Lambda$CDM cosmology. The cosmological parameters used are those obtained by The \cite{Plank2014}: the Hubble constant, $H_0=67.1\,{\rm km\,s^{-1}\,Mpc^{-1}}$; the matter density $\Omega_{\rm m} = 0.3175$, the dark energy density $\Omega_{\Lambda} = 0.6824$, the baryon density $\Omega=0.0490$, the power spectrum normalization $\sigma_8=0.8344$ and the power spectrum slope $n=0.9624$.

These simulations include gas cooling via Hydrogen, Helium and various metal lines, Compton cooling and heating via UV background \citep{Haart2012}. Stars form from cold and dense gas particles matching the Kennicutt-Schmidt relation \citep{Kennicutt1998}. The temperature and density thresholds for star formation are $T < 1.5\times10^4\,{\rm K}$ and $n > 10.6\,{\rm cm}^{-3}$ respectively.
Newly formed stars inject thermal energy into their surroundings. This early energy injection, usually referred to as "Early Stellar Feedback" (ESF), is meant to represent the stellar winds and the ionization from the bright young stars. In NIHAO this ESF injects a $\epsilon_{\rm ESF}=13\%$  of the total stellar flux ($2\times10^{50}\,{\rm erg\, M_{\odot}^{-1}}$).
Regarding supernova feedback, massive stars inject energy and metals into their surroundings. The supernovae feedback is implemented via the blastwave formalism from \cite{Stinson2006}. Cooling is delayed inside the blast region by $30\,{\rm Myr}$, in order to prevent the very efficient cooling from high density gas. Metals injected in the inter-stellar medium by the supernovae are allowed to move between gas particles following the metal diffusion scheme from  \cite{Wadsley2008}.

The resolution of the zoom-in region has been set to properly resolve the mass profile of the central galaxy down to $1\%$ of the virial radius, leading  thus to variable force softening lengths. This way each halo contains approximately $\sim10^6$ particles. The dark matter(gas) force softening lengths are $200(89)~{\rm pc}$ in the less massive ones and $470(200)~{\rm pc}$ in the most massive. 

In this analysis we will focus on Ultra-Diffuse galaxies. For selecting UDGs we follow the prescription of \cite{DiCintio2017}, later on extended in \cite{CardonaBarrero2020}: we  selected all  galaxies with face-on cylindrical half light radius (${\rm R}_{1/2}^{r}$) larger than $1$~kpc and effective surface brightness larger than $\left<\mu_{\rm eff}\right> > 23.5 {\rm mag\, arcsec}^{-2}$, both computed in the $r$-band. To compute the effective surface brightness we use the following conversion:
\begin{equation}
    \left<\mu_{\rm eff}\right> = \mathcal{M}_{\odot} + 21.572 - 2.5\log_{10}\left(\frac{\left(L/2\right)/L_{\odot}}{\pi\left({\rm R}_{\rm 1/2}^{r}/{\rm pc}\right)^2}\right), 
\end{equation}
being $\mathcal{M}_{\odot}$ the sun's absolute magnitude and $L$ the total luminosity of the galaxy, both in the $r$-band. 
This selection  leads to a sample of $35$ ultra diffuse galaxies in the stellar mass range between $6\times10^6\,{\rm M}_{\odot}$ and $10^9\,{\rm M}_{\odot}$.

\section{Results}
\label{sec:results}

\begin{figure*}
    \centering   
    \includegraphics[width=\textwidth ]{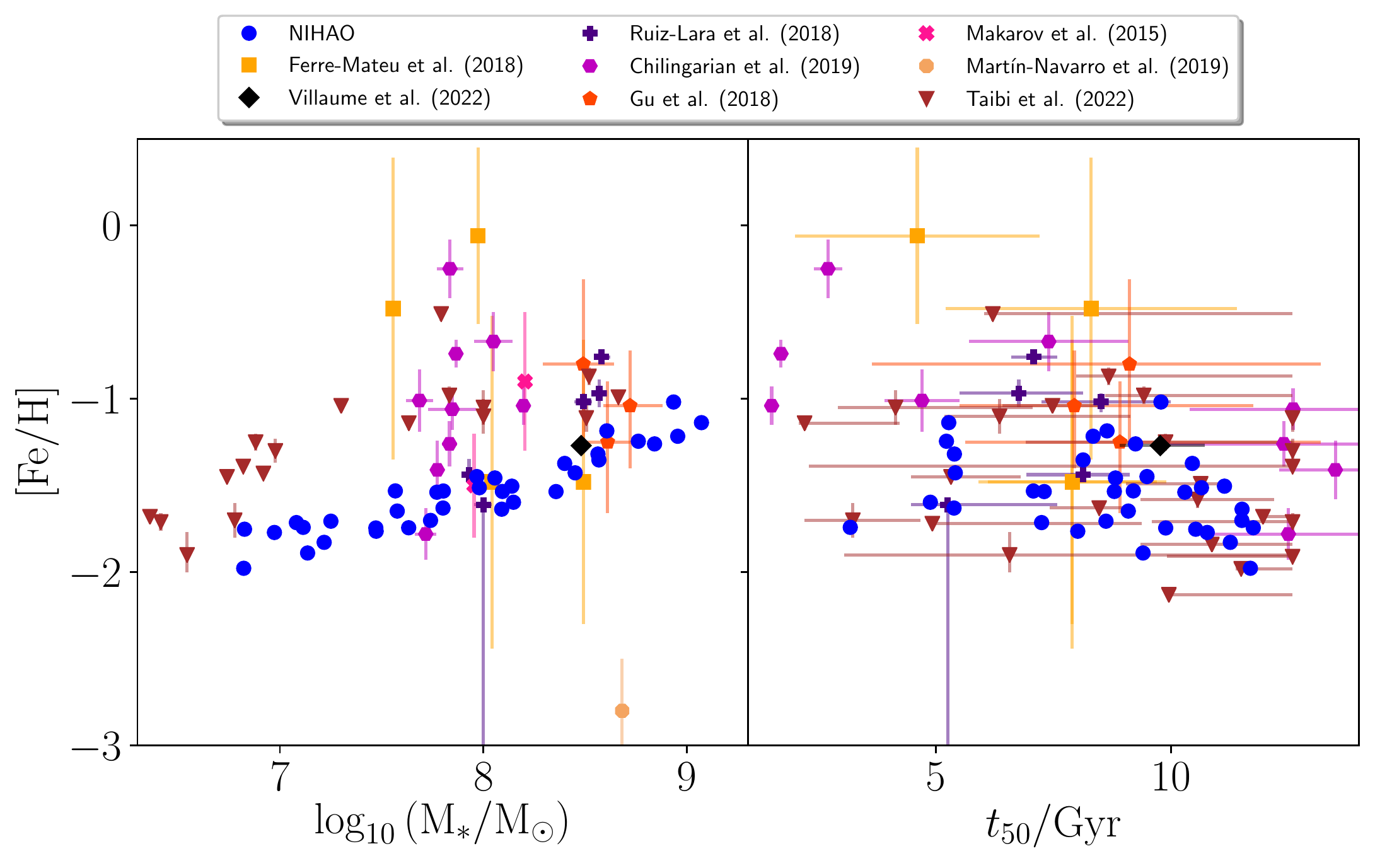}
    \caption{Stellar Mass-Metallicity (left) and Age-Metallicity (right) relation for NIHAO simulated UDGs (blue) and observed UDGs. The observed UDGs are from: \citealt{Villaume2022} (black diamond), \citealt{Chilingarian2019} (purple hexagon), \citealt{Makarov2015} (pink cross), \citealt{FerreMateu2018} (yellow square), \citealt{RuizLara2018} (purple plus symbol), \citealt{Gu2018} (red pentagon) and \citealt{MNavarro2019} (orange octagon). We include as a comparison, the Stellar Mass-Metallicity and Age-Metallicity relations of Local Group dwarfs (brown inverted triangles) form \protect\cite{Taibi}.
    NIHAO UDGs seem to be metal poorer than observed LG dwarfs while matching the metal poor end of the observed UDGs metallicity distribution (note however that NIHAO UDGs are isolated, while all but the \citet{Martin2019} galaxy are cluster objects). Conversely, the  age-metallicity relation provides a good match between observations and simulations: in there,
    the age of NIHAO UDGs is the median age of the stars ($t_{50}$).  
    }
    \label{fig:Mass-Met_rel}
\end{figure*}

In this section we show that the metallicities of simulated NIHAO UDGs, both their average values as well as  their  radial profiles, are compatible with observations, by comparing them with a large data set of Local Group dwarfs as well as with some observed UDGs.
We  explore the origin of the metallicity gradients in NIHAO UDGs, and we  compare them with the so-far-sole metallicity gradient measured in DF44 \citep[][]{Villaume2022}, alongside  with a variety of metallicity profiles derived for LG dwarfs \citep[][]{Taibi}.
Finally, we  explore the possible origin of such gradients, and how are they related to the formation of  UDGs in simulations.

\subsection{UDGs average stellar metallicities}
\label{subsec:metalls}

Before showing the mass metallicity relation of simulated versus observed UDGs, some general considerations must be done.
Comparing observed and simulated metallicities should be done with care, since the exact way of computing the average metallicity can provide completely different results \citep[see, as an   example,][]{Arora2021}.
Computing the average in linear metallicities, $\overline{[{\rm Fe/H}]}_{\rm lin} $, leads in general to larger values than using the metallicities in logarithmic scale $\overline{[{\rm Fe/H}]}_{\rm log}$, which are more sensitive to the low metallicity tail, while the opposite happens in linear average metallicities. 
Here, the linear and log metallicities are defined as:
\begin{align}
    \overline{[{\rm Fe/H}]}_{\rm lin} &= \log_{10}\left(\sum_i \omega_i 10^{[{\rm Fe/H}]_i}  \right), \\
    \overline{[{\rm Fe/H}]}_{\rm log} &=\sum_i \omega_i [{\rm Fe/H}]_i,
\end{align}
with $[{\rm Fe/H}]_i$ being the iron abundance of the $i-$th particle, and $\omega_{i}$ the weights defined in such a way that $\sum_i\omega_i = 1$.

Our tests suggest that, in general, the discrepancy between these two values seems to be of the order of $\sim0.2$~dex, but it can increase up to $\sim0.5$~dex in the mass range we are working in. 

The metallicity estimation is also sensitive to the weights used in the averaging process. Mass weighted and flux weighted metallicities probe different stellar populations, being the flux weighted averages more biased towards the younger stellar populations. Our tests show a median systematic difference between mass weighted and flux weighted metallicities of $-0.1$~dex.

In general, the most appropriate estimator of the metallicity depends on the purpose of the metallicity estimation \citep[see for example ][]{Genina2019}. In order to overcome these issues, we decide to provide median metallicities which are robust to monotonic transformations in the data-set such as the logarithm, and no weighting is required. Another possible source of systematic effects is the existence of metallicity gradients. Our tests show that using the full extent of the galaxy yields, in average, median metallicities $\sim0.05$~dex  lower than when using only the star particles within the $3$D half mass radius. Due to this small systematic effect, which is well below the errors of observed metallicities in UDGs, we decided to use the full extent of the galaxy to measure its global metallicity. 

In the left panel of Fig.~\ref{fig:Mass-Met_rel}, we show the Stellar Mass-Metallicity relation of NIHAO UDGs as blue points, the one of Local Group dwarfs from  \cite{Taibi} as brown downward triangles, and the one of observed UDGs with several symbols, as indicated in the legend \citep[][]{Villaume2022, Chilingarian2019, Makarov2015, FerreMateu2018, RuizLara2018, Gu2018}. For  UDGs observations, we decide to  show only those metallicities obtained from spectroscopic data (but see for example \citealt{Barbosa2020} who provides metallicities and ages for a large sample of UDGs via SED fitting). 
When needed, we transformed total metallicities [M/H] to iron abundance [Fe/H] using the following relation from \cite{Salaris2005} (using [Mg/Fe] as a proxy for the alpha enhancement):
\begin{equation}
    [{\rm M/H}] = [{\rm Fe/H}] + \log_{10}\left(0.694\times10^{[\alpha{\rm /Fe}]} + 0.306\right),
\end{equation}

NIHAO average mass-metallicity relation, and consequently NIHAO UDGs, seem to lie slightly below the LG mass-metallicity relation \citep[see][]{Buck2021}. This is likely due to the explosive feedback that expels too many metals from the galaxies \citep[see][for a detailed discussion about the difficulties of matching the mass-metallicity relation]{Agertz2020, Buck2021}.\\
When comparing with observed UDGs with measured metallicities, NIHAO UDGs provide a good match, although being mostly in the low metallicity regime. 
However, it must be taken into account that  most of these observed UDGs  have evolved in dense environments such as the Coma cluster, which can completely affect their actual metallicity. In fact, \cite{Sales2020}, using Illustris-TNG, showed that tidally stripped UDGs are likely to have larger metallicities than the expected from their stellar mass. NIHAO UDGs are isolated galaxies, thus a systematic difference in their metal content may be expected when comparing with in-cluster UDGs.
We conclude that despite the difficulties of properly comparing metallicities between simulated systems and observations, isolated NIHAO UDGs seem to match the low metallicity tail of observed cluster UDGs, while being  slightly more metal poor than same mass LG dwarfs, residing in a group.

\subsection{UDGs average stellar ages}

Similar systematic effects as the ones described above are involved in the computation of stellar ages. The median offset between flux and mass-weighted mean ages is $\sim2$ Gyrs. Following the same argumentation as in Section~\ref{subsec:metalls}. we decided to measure the ages of the NIHAO UDGs as the median age of the stars at $z=0$ ($t_{50}$).

Regarding the ages, NIHAO UDGs tend to be systems with old and intermediate median ages, as shown in the right panel of  Fig.~\ref{fig:Mass-Met_rel}), none of them having a median age ($t_{50}$) younger than  $3$~Gyrs ($t_{50}<3\,{\rm Gyrs}$). Despite of this, NIHAO UDGs show rather extended star formation histories, 
with residual star formation up to $z=0$. 
We do find a reasonable match between the median ages of NIHAO UDGs and observed UDGs, with differences  well below the mean uncertainty in the observed ages and of the order of the possible systematics we have found when measuring ages in simulated systems.

As opposite to the tight mass-metallicity relation shown in the left panel of Fig.~\ref{fig:Mass-Met_rel}, we do not find a strong correlation between  median ages and iron-abundances, neither in observations nor in simulations, indicating that  ages should  be used with care as proxy for  metallicity in these galaxies, and viceversa. This behavior also appears in  LG-dwarfs.

\subsection{Metallicity Gradients}
\subsubsection{Description of radial profiles}

\begin{figure*}
    \centering
    \includegraphics[width=.65\textwidth]{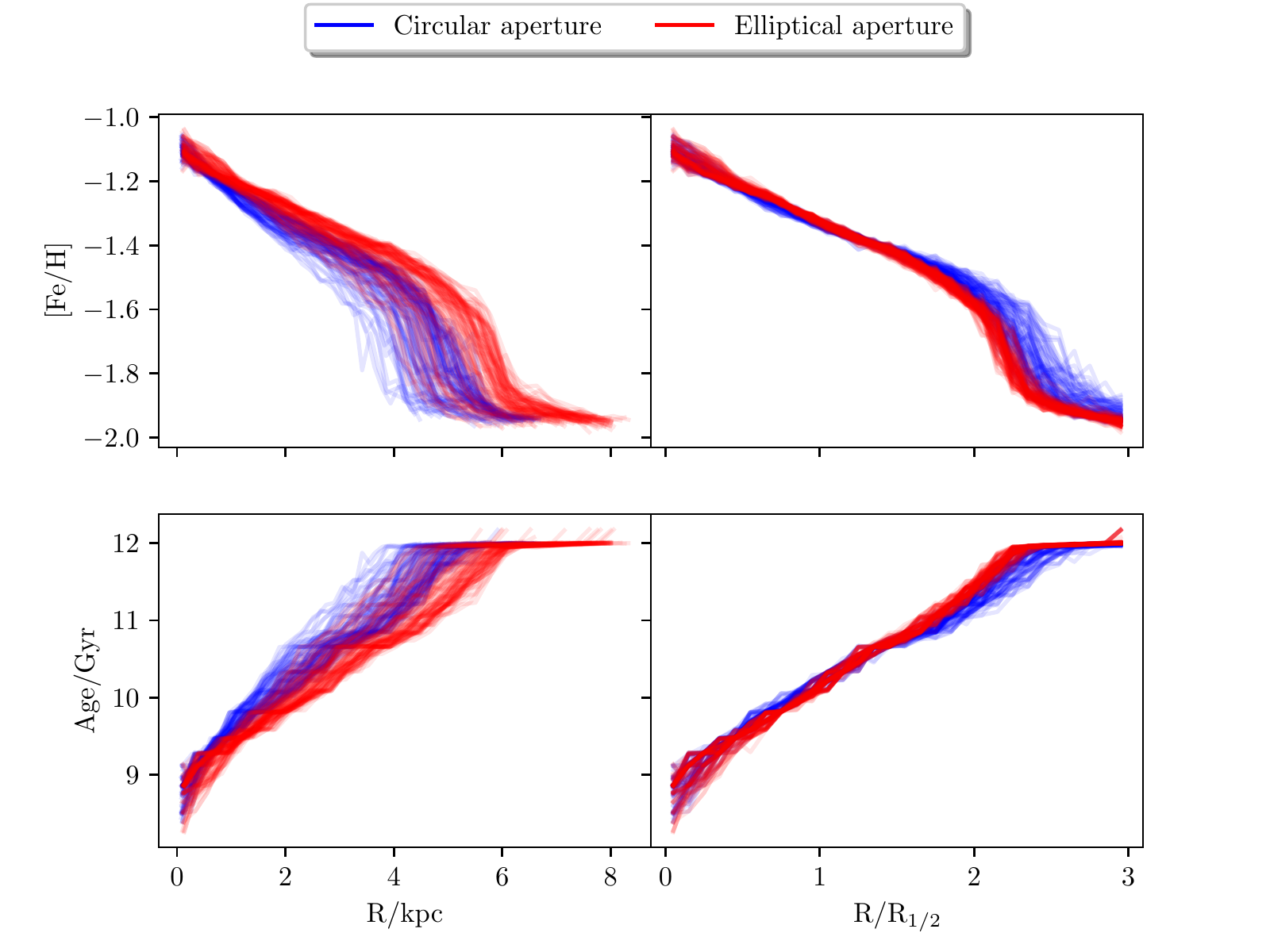} 
    \includegraphics[width=.345\textwidth]{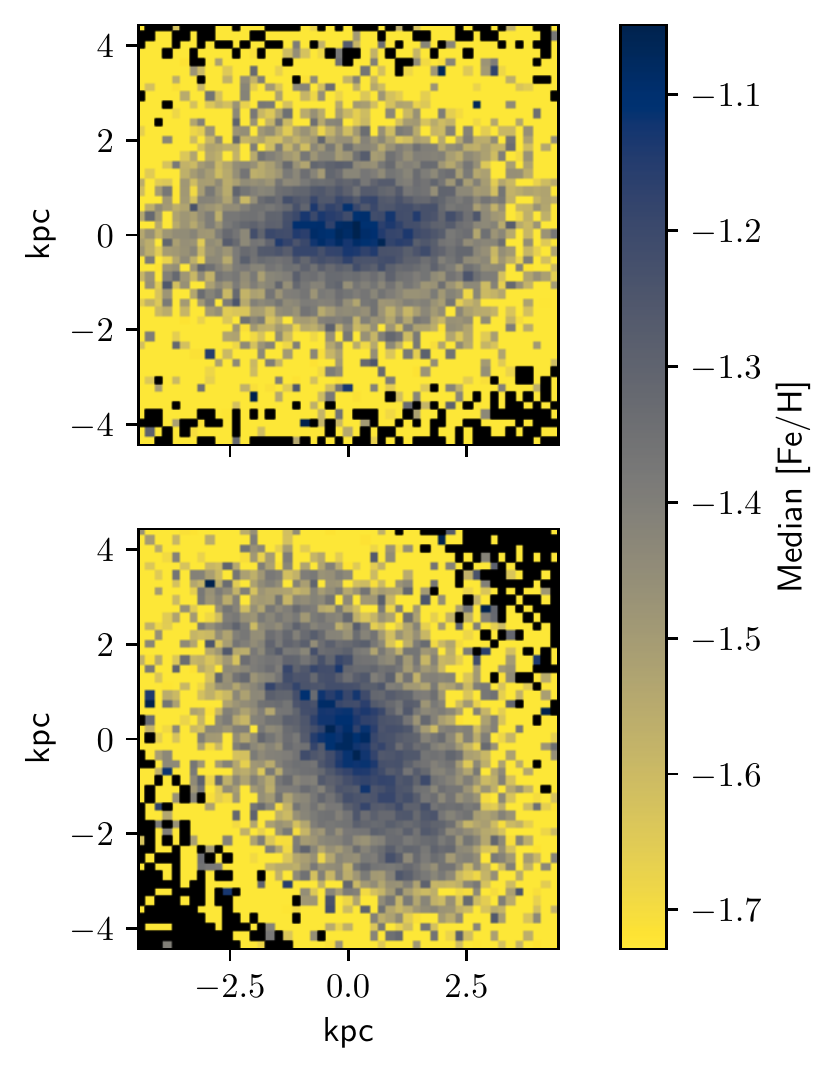}
    \caption{Example of stellar metallicity gradients. 
    The left and central panels show  metallicity (top) and the age (bottom) profiles constructed for one of the NIHAO UDGs in physical units and normalized by the stellar half-mass radii respectively.
    Each line represents one of $100$ random projections. Blue lines are the profiles built using circular annuli while red lines show the profiles using elliptical annuli. The variations in the profile due to projection effect are minimized when using coordinates normalized to the half-mass radius. 
    Right panels show the $2$D median metallicity maps of the same galaxy in $2$ different random orientations. This maps show that, due to the morphology of the galaxy, elliptical annuli better describe the stellar radial metallicity distribution.}
    \label{fig:MetProfilesExm}
\end{figure*}

To characterize the metallicity profile of NIHAO UDGs, we constructed $2$D metallicity and age radial profiles. In order to avoid possible projection effects, for each galaxy we have constructed $100$ different projections in randomly chosen orientations. Each profile has been derived up to $3$ times the projected half mass radius of the stellar component (${\rm R}_{1/2}$) using bin widths of $0.1{\rm R}_{1/2}$. Each bin-value corresponds to the median value of the particles inside the bin.

In Fig.~\ref{fig:MetProfilesExm} we show the metallicity and age profiles of one of the simulated UDGs. The left panels show the profiles in physical coordinates while the central panels show the profiles normalized by ${\rm R}_{1/2}$. 
We decided to use the profiles normalized by ${\rm R}_{1/2}$ to characterize the metallicity gradients as this normalization seems to reduce the variability due to projection effects. This normalization is a common practice in works analyzing age and metallicity gradients \citep[e.g. ][]{Mercado2021}.
Another source of variability while constructing this type of radial profiles is whether to use Elliptical (red) or Circular (blue) annuli. In the right panels of Fig.~\ref{fig:MetProfilesExm} we show  examples of median iron abundance maps, for a same galaxy but in  different projections, indicating that the radial variation of the median metallicity is better described by considering elliptical annuli: this is what we will adopt hereafter \footnote{Note that for the elliptical annuli we use distances along the projected semi-major axis.}. 

\begin{figure}
    \centering
    \includegraphics[width=1\columnwidth]{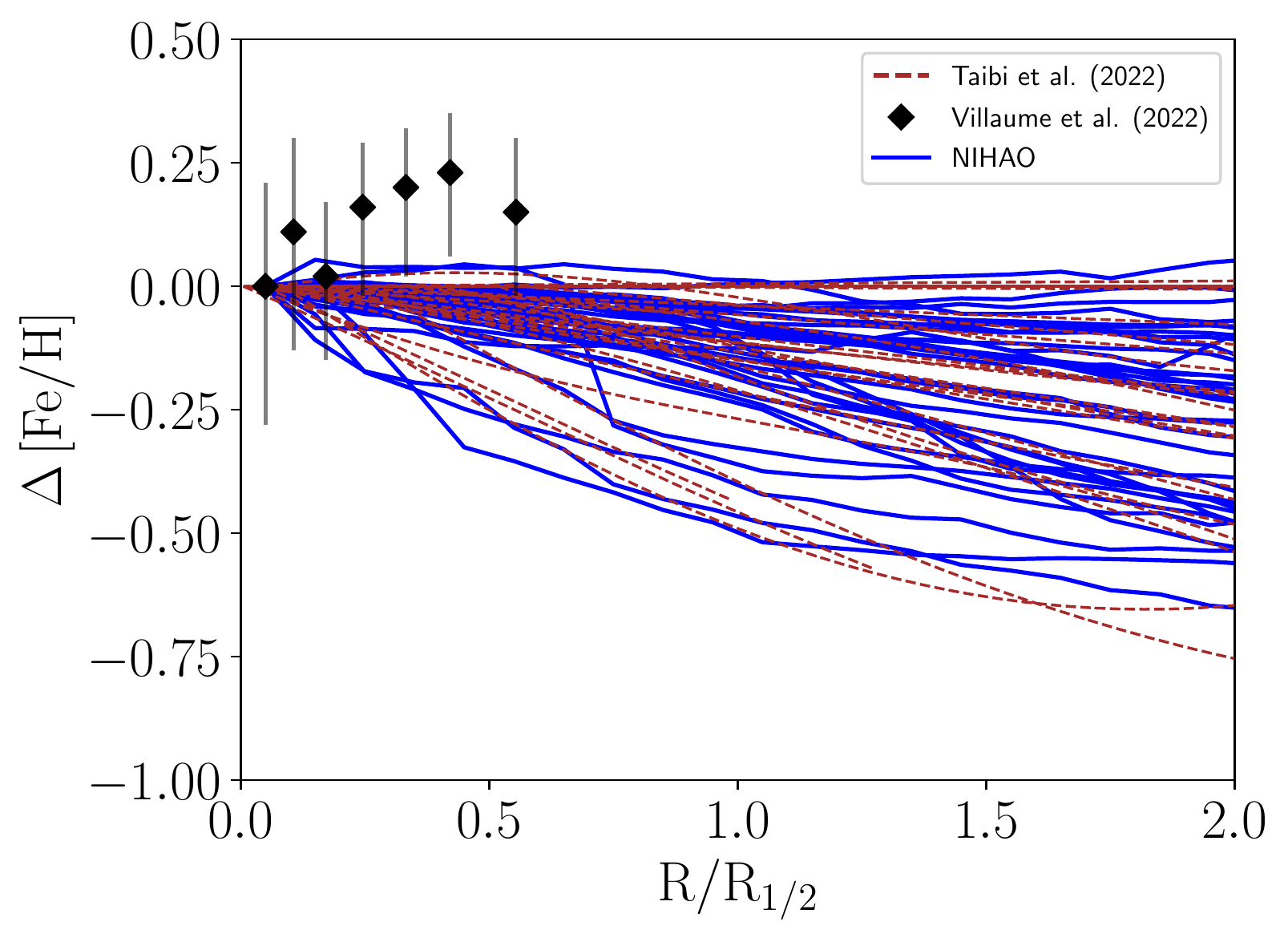}
    \caption{Stellar metallicity profiles of NIHAO UDGs (blue), LG dwarfs \protect\citep[dashed red, ][]{Taibi}  and observed UDG DF$44$ \protect\citep[black diamonds, ][]{Villaume2022}. The profiles have been normalized to their central value for easier comparison. Simulated UDGs and LG dwarfs show similar gradients in stellar metallicity, while the so-far-only-measured
    metallicity profile of an observed UDG shows a positive gradient.}
    \label{fig:MetProfilesStacked}
\end{figure}

The metallicity profiles for all the NIHAO UDGs can be found in Fig.~\ref{fig:MetProfilesStacked}. For easier visualization for each galaxy we plot the median profile over the $100$ orientations. All the profiles have been normalized to their central values in order to highlight the large variety of shapes presented in the sample.
Metallicity profiles of NIHAO UDGs are in general  flat to negative, with some examples of steep negative metallicity gradients. We only find two examples of systems in which the metallicity increases with radius, and it does it smoothly only in the very inner regions. 
In Fig.~\ref{fig:MetProfilesStacked} we also include the metallicity profiles of the LG dwarfs (red dashed lines) from \cite{Taibi}, indicating a very good agreement with our simulated  UDGs, in terms of diversity and steepness of the profiles.

We have also included in Fig.~\ref{fig:MetProfilesStacked} the recent measurement of a metallicity profile of  UDG DF$44$ made by \cite{Villaume2022}. We do not find any NIHAO UDG with a positive metallicity gradient as the one found in DF$44$.
Note, however, the very large error bars and the small radial range covered by DF$44$. Nevertheless, we do find two systems with shallow increasing metallicity profiles (up to $1{\rm R}_{1/2}$) which show quantitative different behaviours: a plateau of metal rich population that quickly drops around $\sim0.7{\rm R}_{1/2}$ and a steady shallow increase of the metallicity up to  $1{\rm R}_{1/2}$ which decreases smoothly afterwards. These two systems are characterized by strong bursts of recent star formation triggered by late time gas accretion.  The different morphology of the metallicity  profiles of these two galaxies is related to the time scales in which the late time SF happens.

In the first galaxy, the central plateau of high metallicity is related to a very recent SF event in the inner regions, which represents $\sim23.8\%$ of the galaxy stellar mass at $z=0$. These newly formed stars do not have enough time to mix with the older ones, thus there is a sharp transition at $\sim0.7{\rm R}_{1/2}$ in which the SF is confined. On the other hand, the shallow increasing metallicity of the second UDG seem to be related to $3$ different bursts of SF triggered by gas accretion of pre-enriched material. The larger time scales involved allow a more efficient mixing of the populations avoiding discontinuities in the profiles.

This is not likely the case of DF$44$ as no hints of recent SF have been found in that galaxy \citep{Villaume2022}. More data of observed metallicity gradients in UDGs are needed, in order to confirm or discard our prediction.
For the time being, we conclude this section by noting that simulated UDGs seem not to be peculiar in what regards their metallicity gradients, when compared to regular dwarf galaxies from the LG.
Furthermore, our analysis highlights the importance of reaching larger distances  in order to capture the full shape of the metallicity profile of such elusive galaxies.

\subsubsection{NIHAO UDGs: feedback driven metallicity gradients? }

\begin{figure*}
    \centering
    \includegraphics[width=.8\textwidth]{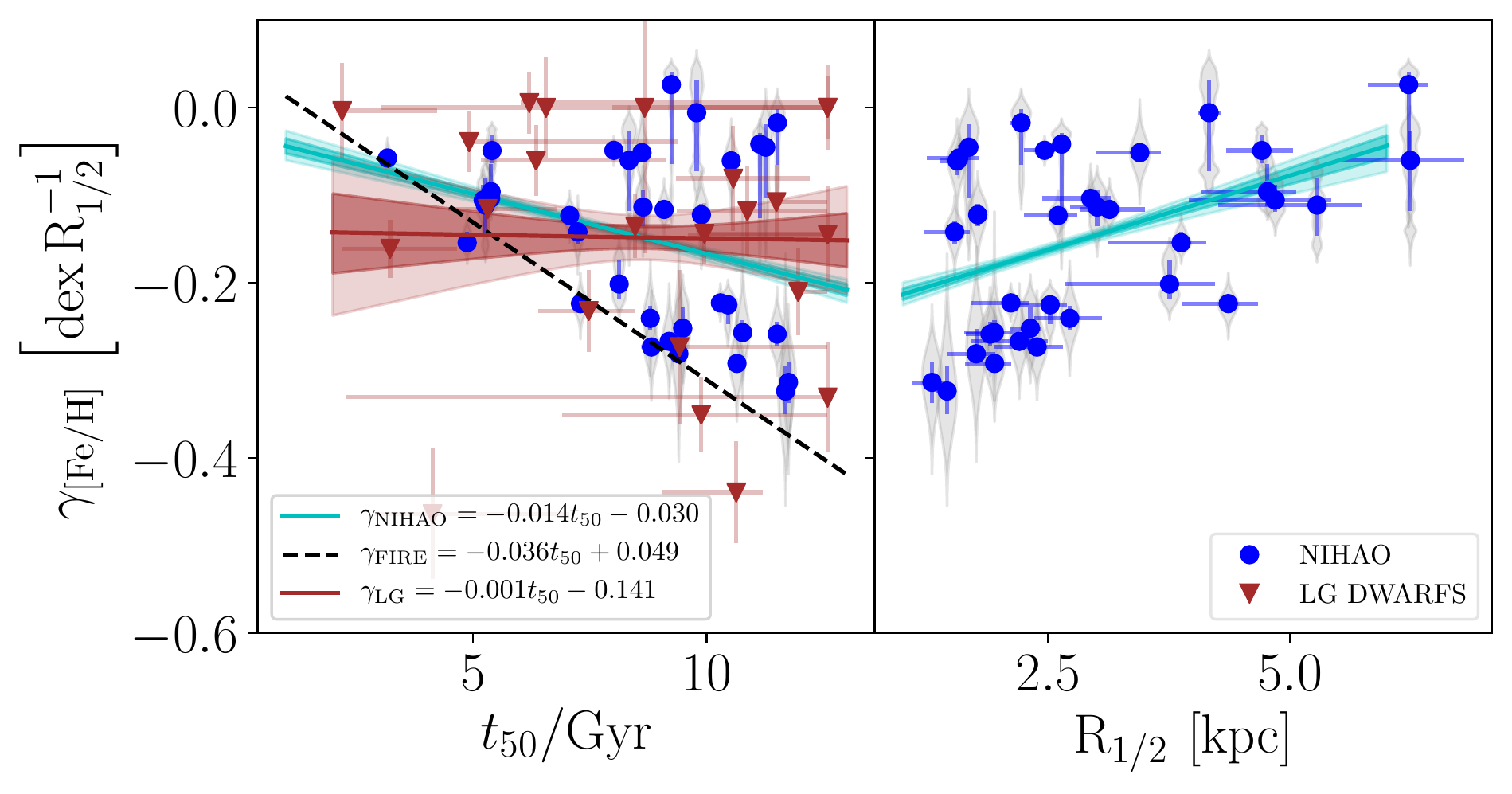}
    \includegraphics[width=.8\textwidth]{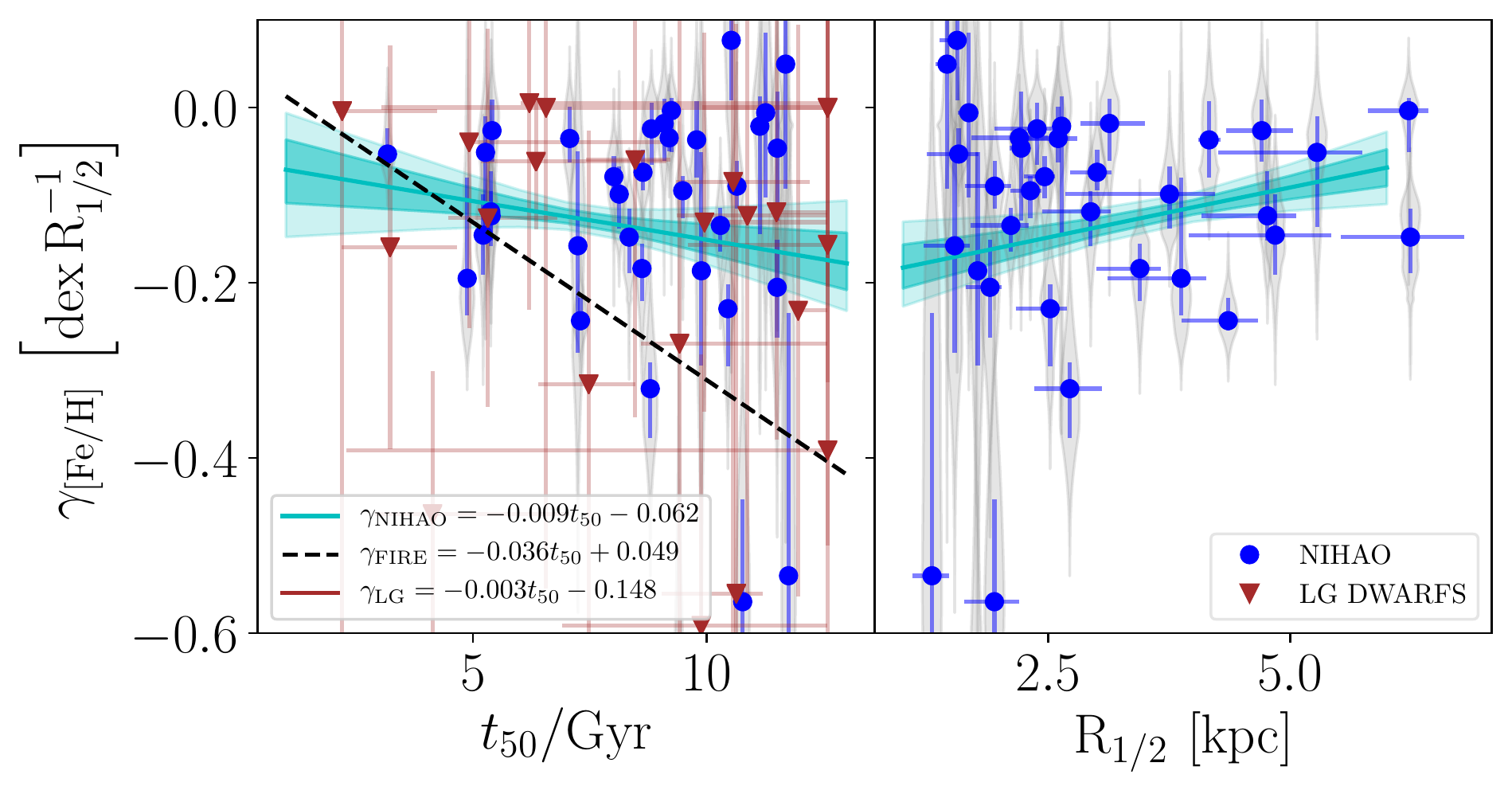}
    \caption{
    Metallicity slope $\gamma_{\rm [Fe/H]}$ versus median age $t_{50}$ (left) and projected half-mass radius ${\rm R}_{1/2}$ (right). We show the effects of changing the aperture by displaying the metallicity gradients measured inside $2{\rm R}_{1/2}$ (top) and $0.6{\rm R}_{1/2}$ (bottom).
    NIHAO UDGs are shown as blue circles. The shadowed grey area around each point shows the metallicity slope distribution for each UDG obtained from $100$ random orientations. The horizontal error-bar in the right panels indicates the range of projected half-mass semi-major axis radii measured in the different projections. 
    We also include data for the LG dwarfs (brown downwards triangles): the metallicity gradients have been measured from the profiles of \protect\cite{Taibi} in the same aperture used for NIHAO UDGs. Median ages from the LG dwarfs have been obtained from the published SFH in \protect\cite{Weisz2014}. 
    The black dashed line shows the linear relation between the metallicity gradient and the age obtained by \protect\cite{Mercado2021} using FIRE-2 simulations.
    Cyan and brown lines show the linear fits obtained for the NIHAO UDGs and the LG dwarfs respectively (see the text for details). In both cases the line shows the median fit and the shaded regions the inter quantile rages: $2.5{\rm th}-97.5{\rm th}$ and $16{\rm th}-84{\rm th}$. NIHAO UDGs and LG dwarfs both suggest a very weak, if any at all, simple linear correlation between stellar metallicity gradients and median stellar ages, and NIHAO data further suggest a small correlation between $\gamma_{\rm [Fe/H]}$ and the galaxy size.
     }
    \label{fig:SlopeAge_rel}
\end{figure*}

In order to quantitatively characterize the metallicity gradient $(\gamma_{\rm [Fe/H]})$ of NIHAO UDGs we perform a linear fit to each of the metallicity profiles described in the previous section, having thus $100$ gradients per galaxy (one per projection). We found that the exact value of the metallicity gradient is strongly dependent on the aperture used to measure it. This can be qualitatively seen in Fig.~\ref{fig:MetProfilesStacked}  and quantitatively in Fig.~\ref{fig:SlopeAge_rel}. In general, a linear fit for the metallicity  may not be fully appropriate, however seems to be enough to provide insights about the evolution of individual systems \citep[e.g.][]{BenitezLlambay2016,ElBadry2016,Revaz2018,Mercado2021,Villaume2022}. 
We decided thus to compute the average metallicity gradients considering a region of $2{\rm R}_{1/2}$ from the galaxy center.

\cite{Mercado2021}, using a sample of $26$ dwarfs with masses within $10^{5.5}$\Msun\ and $10^{8.6}$\Msun\,  from FIRE-2  simulations \citep[][]{Hopkins2018}, found a strong correlation between the galaxy age, measured as the median age of the stellar component, i.e. $t_{50}$, and its metallicity gradient. They interpreted this strong relation as the interplay between the migration of old stars towards the outskirts of a galaxy, which steepens the metallicity gradient, and a late time gas accretion, which make its flatter. Moreover, they claimed to have found a similar relation using  LG dwarfs \citep[see, however, the recent work of][in which such a relation is not reported.]{Taibi} 

In the framework of NIHAO simulations, UDGs are systems whose structural properties are strongly dominated by stellar feedback and outflows \citep[][]{DiCintio2017}, consequently they are the perfect laboratory to explore the relation proposed by \cite{Mercado2021}. In Fig.~\ref{fig:SlopeAge_rel} we show the metallicity gradients versus the median age of the stars $t_{50}$ (left) and the half-mass radii ${\rm R}_{1/2}$ (right). We show the gradient measured in an aperture of $2{\rm R}_{1/2}$ (top) and $0.6{\rm R}_{1/2}$ (bottom) in order to highlight the aperture effects in the measurements of the gradients.
Indicated for each NIHAO galaxy is its median value of $\gamma_{\rm [Fe/H]}$ and associated error as blue points with error bars.  On this plane, NIHAO UDGs appear to follow two loci, the one occupying the lower part of the diagram being in general agreement with the relation by \cite{Mercado2021}, indicated as a dashed black line in the left panels of Fig.~\ref{fig:SlopeAge_rel}. On the other hand, the relation by these authors does not capture the behaviour of the rest of the population. 

In order to quantify the relation between $\gamma_{\rm [Fe/H]}$ and $t_{50}$, we performed a linear fit between the metallicity profile and the median age of the stars of the simulated UDGs:
\begin{equation}
\label{eq:lin}
    \gamma_{\rm [Fe/H]} (t_{50}) = \alpha t_{50} + \beta,
\end{equation}
with $\alpha$ and $\beta$ being the slope and y-intercept point, as indicated in the legend within each panel of Fig.~\ref{fig:SlopeAge_rel}.
In order to account for the variability induced by projection effects, we have performed such fit by randomly sampling for each galaxy one value of the metallicity gradient $\gamma_{\rm [Fe/H]}$ from the $100$ projections. We have repeated this procedure $1000$ times.  The best overall relation  is shown as a cyan line in Fig.~\ref{fig:SlopeAge_rel}, while the resulting distribution of  parameters ($\alpha$, $\beta$) are shown in Fig.~\ref{fig:PosteriorFit}. The best fit parameters can be found in Tables \ref{tab:model_params} and \ref{tab:model_params2} for the two different apertures considered in this work: $2{\rm R}_{1/2}$ and $0.6{\rm R}_{1/2}$ respectively.

Our results suggest that, if any at all, the linear relation between the metallicity gradients and the median ages of the UDGs is very weak, being the Spearman correlation coefficient $R_{S}= -0.29$, and not significant, with a P-value $p=0.098$. It is possible that differences in the properties probed by the sample of simulated galaxies are at the origin of the differences seen between this work and that by \cite{Mercado2021}.

On another note, if the metallicity gradients are solely caused by the  stellar redistribution due to explosive SNae feedback, which we know is the responsible for the large sizes of NIHAO UDGs \citep[][]{DiCintio2017}, we would expect a correlation between the metallicity gradient and the size of a galaxy, characterized in terms of the half-mass radii. We explore this possible relation in the right hand panels of Fig. \ref{fig:SlopeAge_rel}. We do find a weak-to-moderate ($R_S=0.42$) and  significant ($p=0.017$) correlation between this two quantities\footnote{Note that the normalization of the metallicity gradients by the stellar half-mass radii, diminishes the strength of the correlation. The same correlation test performed between the metallicity gradient, $\gamma_{\rm[Fe/H]}$, in units of ${\rm dex}\,{\rm kpc}^{-1}$ and the  stellar half mass radius, ${\rm R}_{1/2}$, indicates a stronger ($R_{S}=0.65$) correlation with high significance ($p=5\times10^{-5}$)}.  
This relation is not found in \cite{Mercado2021} dwarfs, highlighting the differences between FIRE and NIHAO models. Our analysis suggest thus that the stellar age, together with  feedback driven stellar redistribution, is not the only driver of the  $z=0$ metallicity gradients in UDGs.

Furthermore, we are  interested in adding context to our results using  LG dwarfs. We obtained the metallicity gradients of the LG dwarfs by performing a linear fit to the radial metallicity profiles of \citet{Taibi} up to $2{\rm R}_{1/2}$ and $0.6{\rm R}_{1/2}$, respectively. The corresponding $t_{50}$ and error-bars have been obtained from the published cumulative SFHs by \cite{Weisz2014}. In order to properly handle the highly asymmetric errors provided by \cite{Weisz2014} we have assumed the error distribution of $t_{50}$ to be described by a beta distribution. We showed LG galaxies as brown downwards triangles with error bars in Fig. \ref{fig:SlopeAge_rel}. The linear fit and confidence region between $t_{50}$ and $\gamma_{\rm [Fe/H]}$  have been obtained by re-sampling the median age and the metallicity slope from the best fitted Beta and Gaussian distributions respectively. The details of the procedure can be found in Appendix~\ref{apx:fits}.

We show the best fitting relation\footnote{We note that when measuring the gradient in apertures of $0.6{\rm R}_{1/2}$ the fit is unconstrained, so we do not show it in the plots.} in Fig. \ref{fig:SlopeAge_rel} together with the $1\sigma$ and $2\sigma$ confidence region. The slope of the obtained relation $\alpha_{\rm LG}$ is compatible with zero within $1\sigma$. Our best fit parameter can be found in Tab.~\ref{tab:model_params} and Tab.~\ref{tab:model_params2}. 
We confirm the findings by \citet{Taibi}: the LG dwarfs do not support a simple  linear relation between the age and the metallicity gradient. Interestingly, this is also the case for the NIHAO UDGs here analyzed. 

\subsubsection{NIHAO UDGs: Alternative origin for the stellar metallicity gradient distribution}\label{subsub:alternatives}

\begin{figure*}
    \centering
    \includegraphics[width=.7\textwidth]{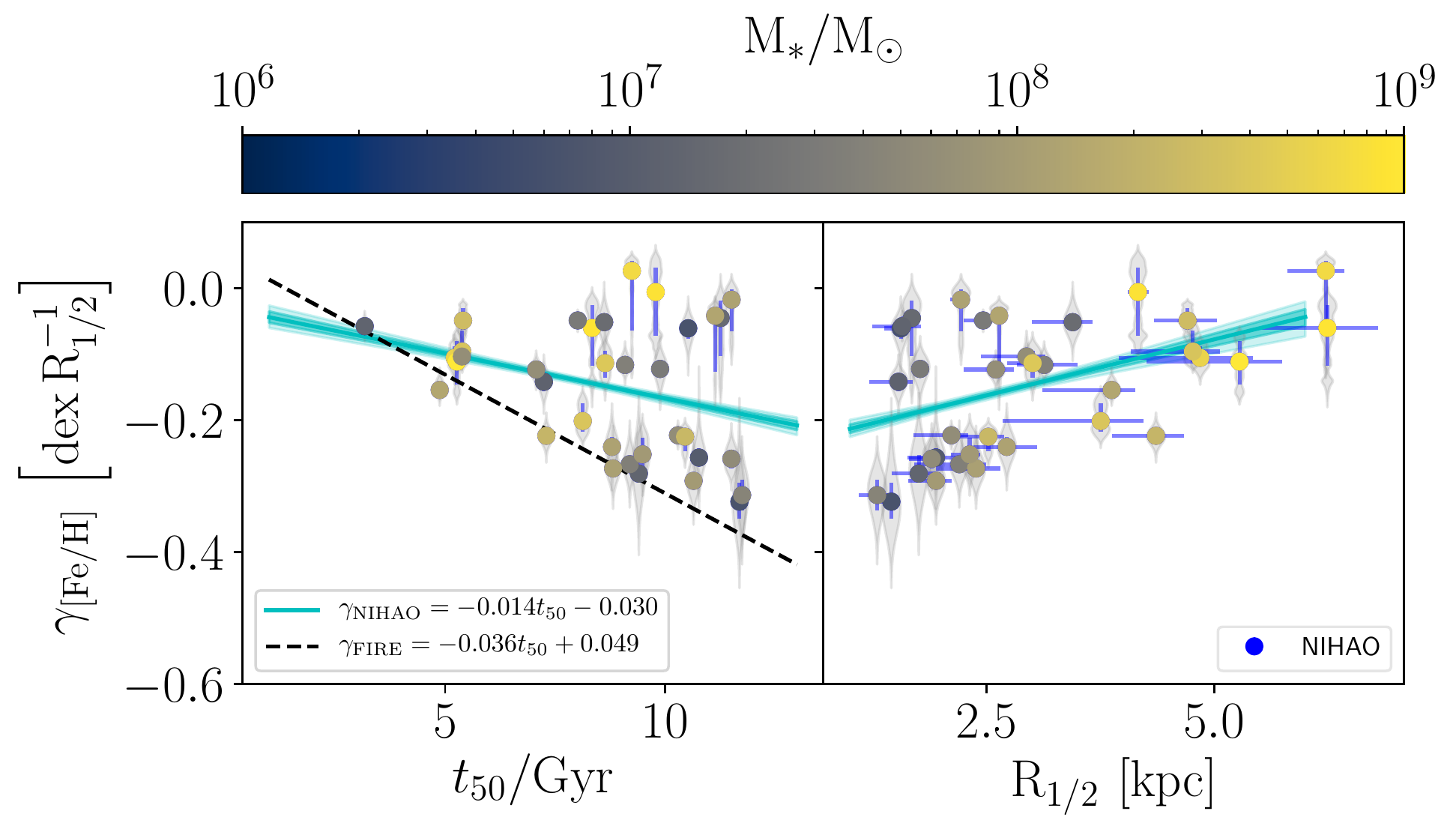}
    \includegraphics[width=.7\textwidth]{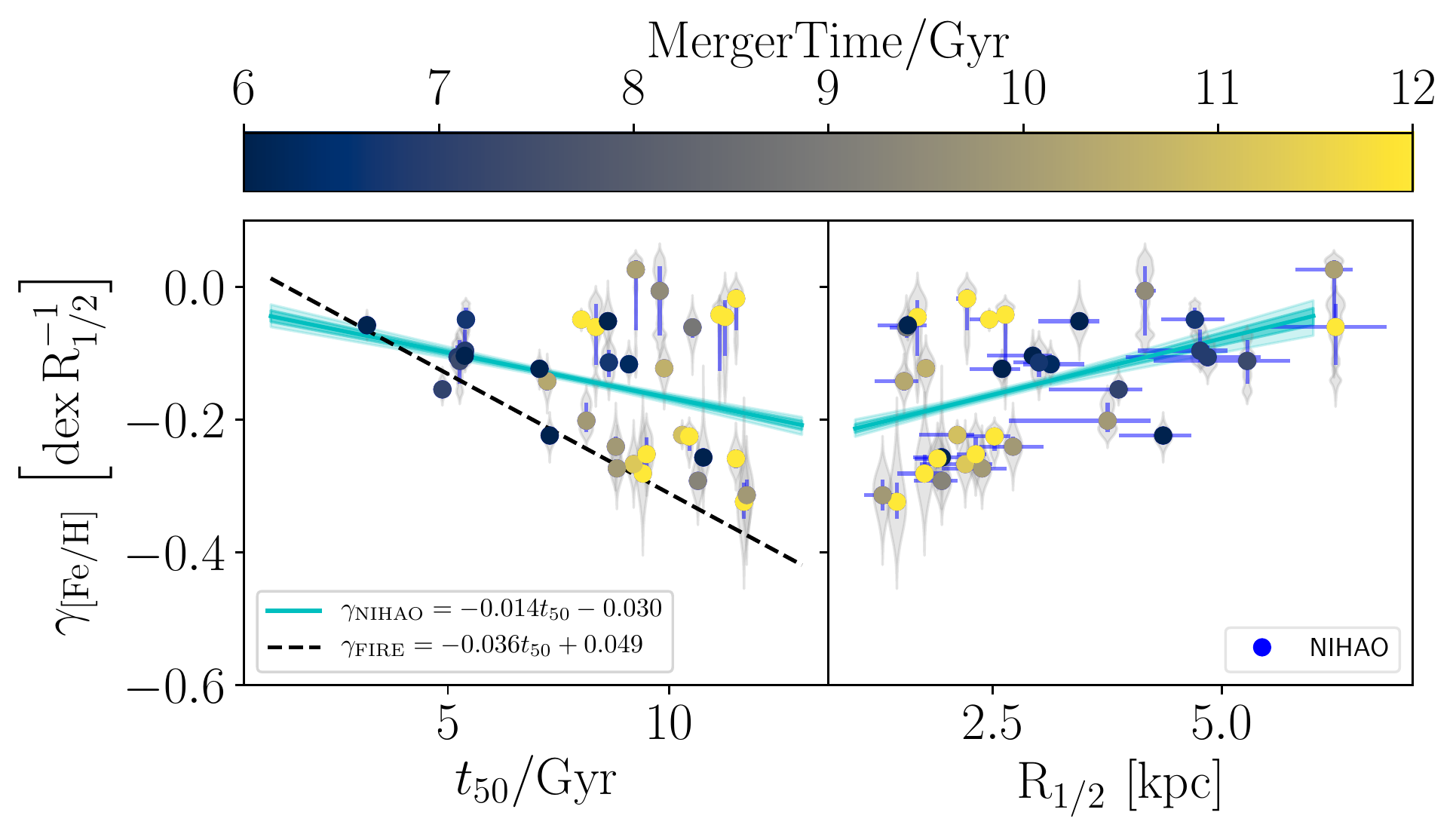} 
    \includegraphics[width=.7\textwidth]{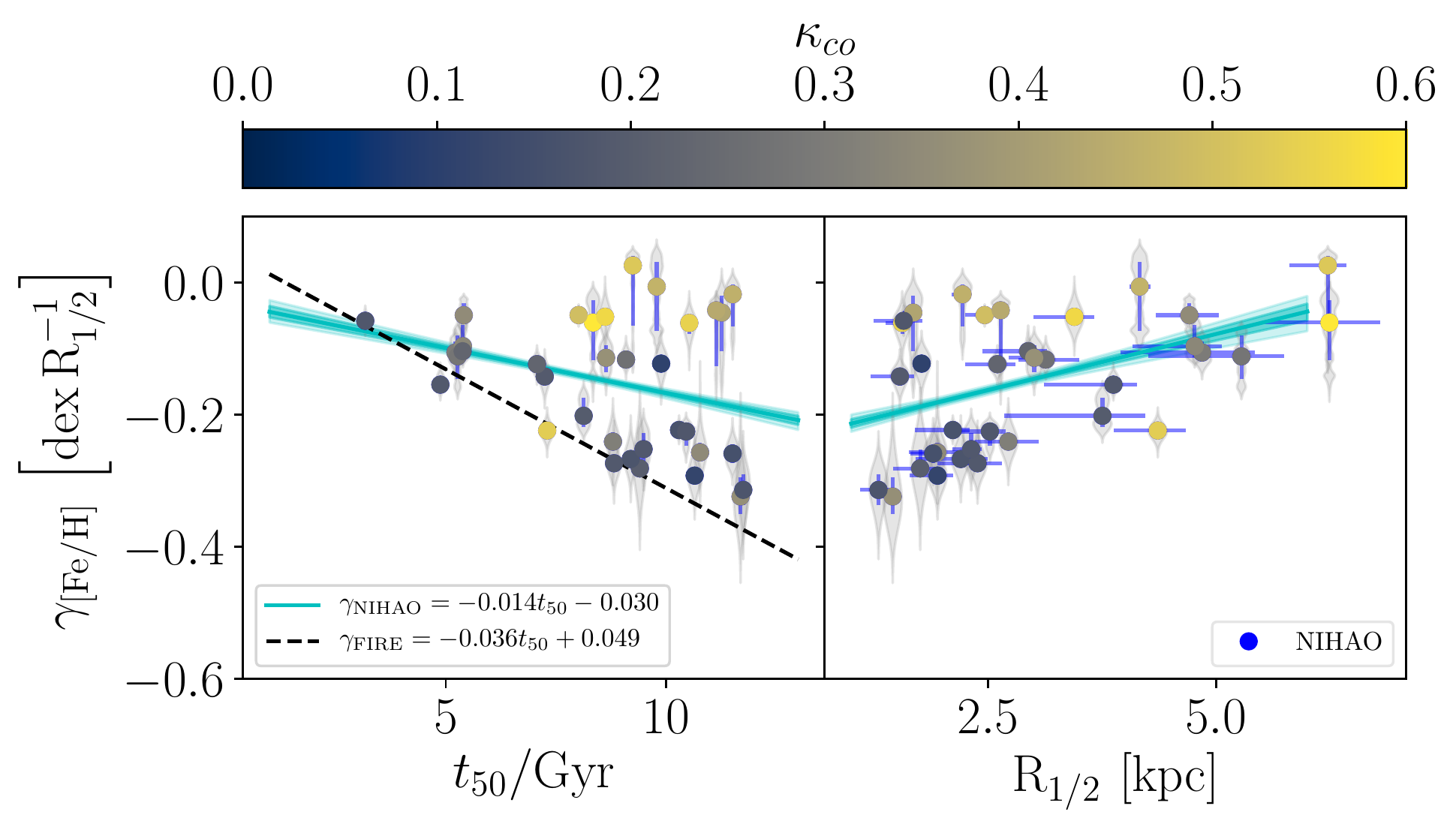} 
    \caption{Metallicity slope $\gamma_{\rm [Fe/H]}$ versus median age $t_{50}$ (left) and projected half-mass radius ${\rm R}_{1/2}$ (right) of NIHAO UDGs. The color-coding indicates stellar mass (top),  time since last major merger (middle) and ordered circular
motion compared to the total kinetic energy of the galaxy  (bottom). The black dashed line shows the linear relation between the metallicity gradient and the age obtained by \protect\cite{Mercado2021}.  The cyan line indicates the linear fit to NIHAO UDGs with the shaded region indicating the inter quantile ranges: $2.5{\rm th}-97.5{\rm th}$ and $16{\rm th}-84{\rm th}$. Rotation supported systems  have on average flatter metallicity gradients  compared to dispersion supported galaxies. Dispersion supported systems seem to follow a tight relation between the metallicity slope and stellar age,  similarly to what reported in  \protect\cite{Mercado2021}. }
    \label{fig:Grad_CCoded}
\end{figure*}

\begin{figure*}
    \centering
    \includegraphics[width=.3\textwidth]{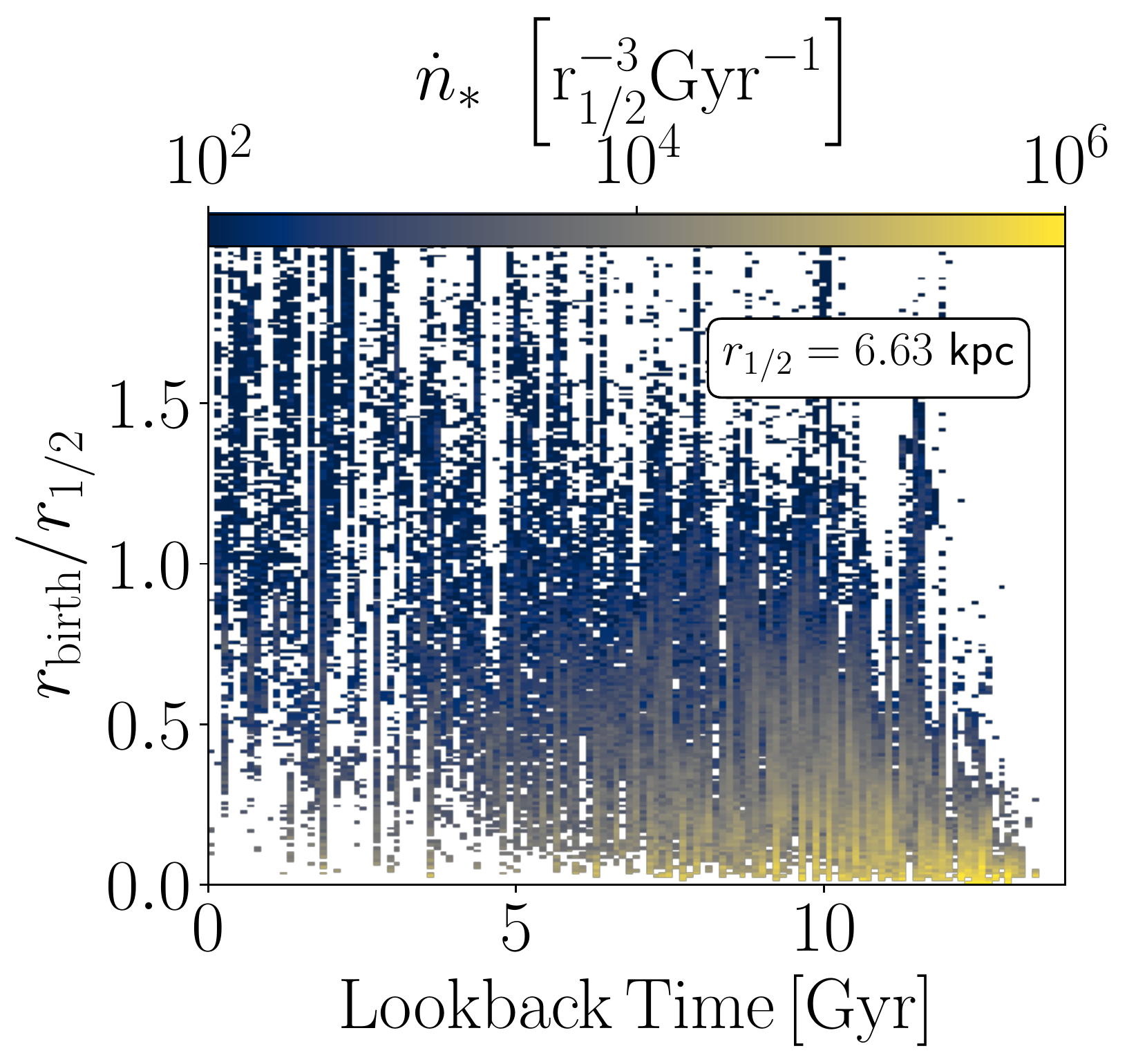}
    \includegraphics[width=.3\textwidth]{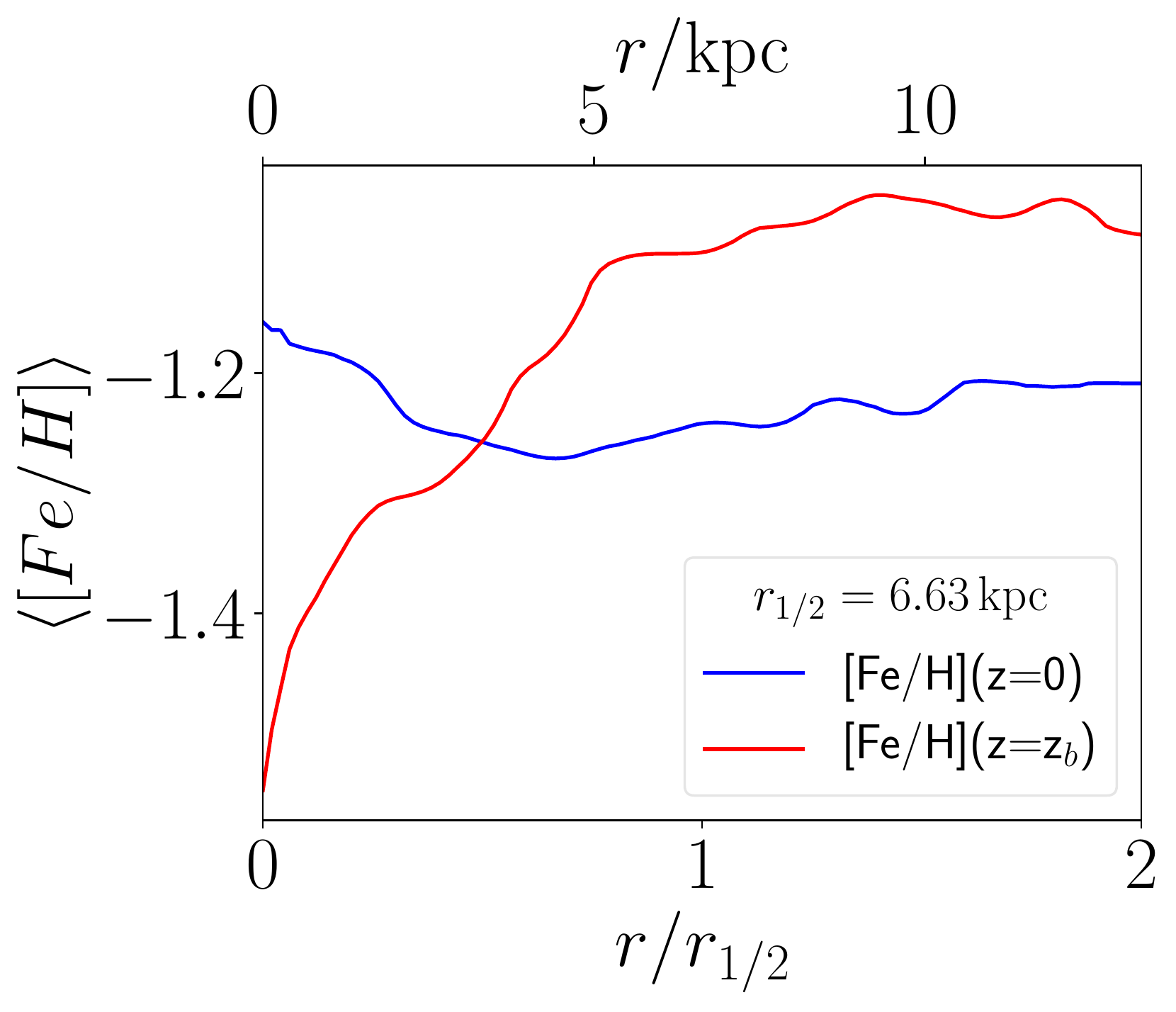}
    \includegraphics[width=.3\textwidth]{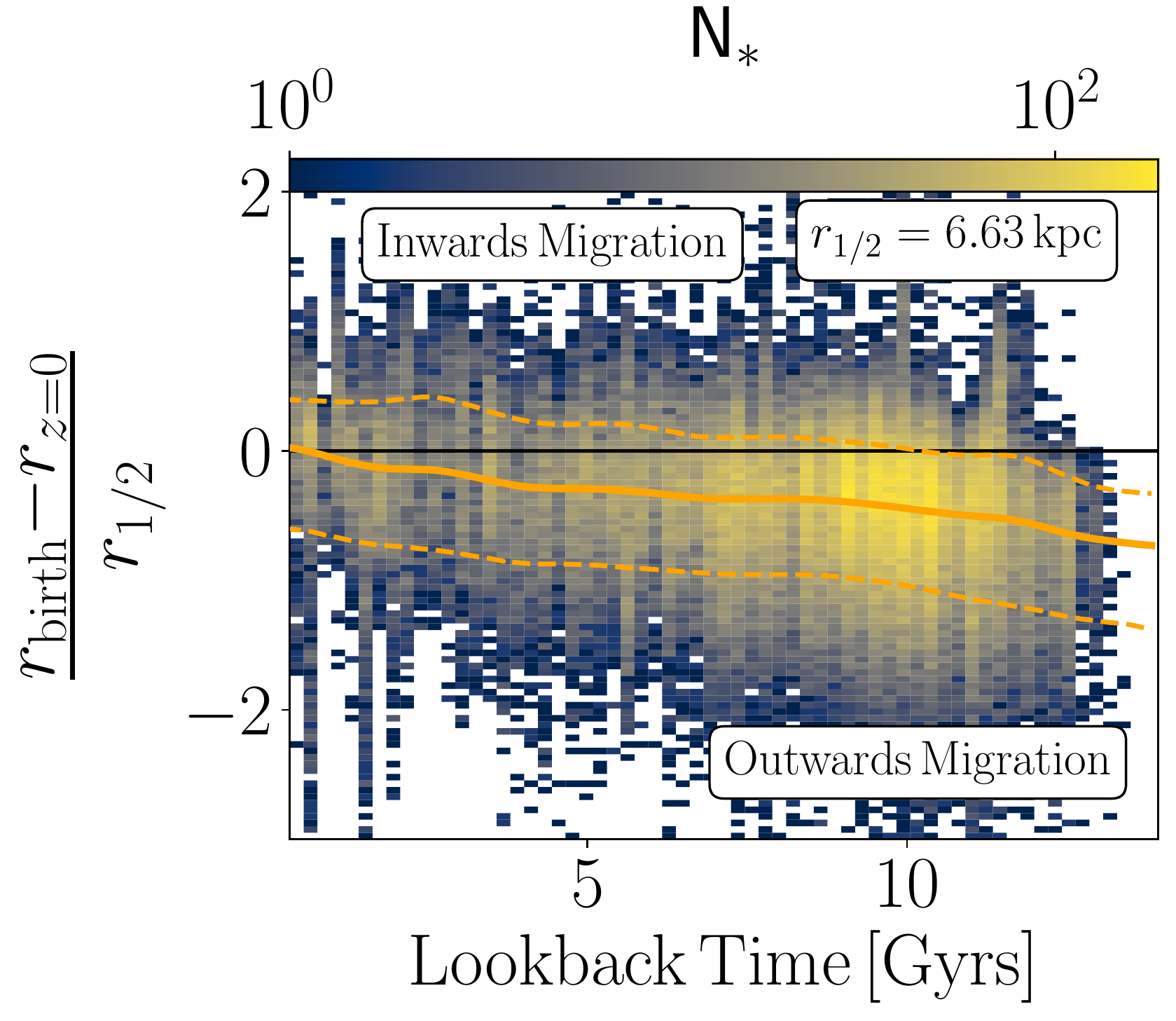}
    \includegraphics[width=.3\textwidth]{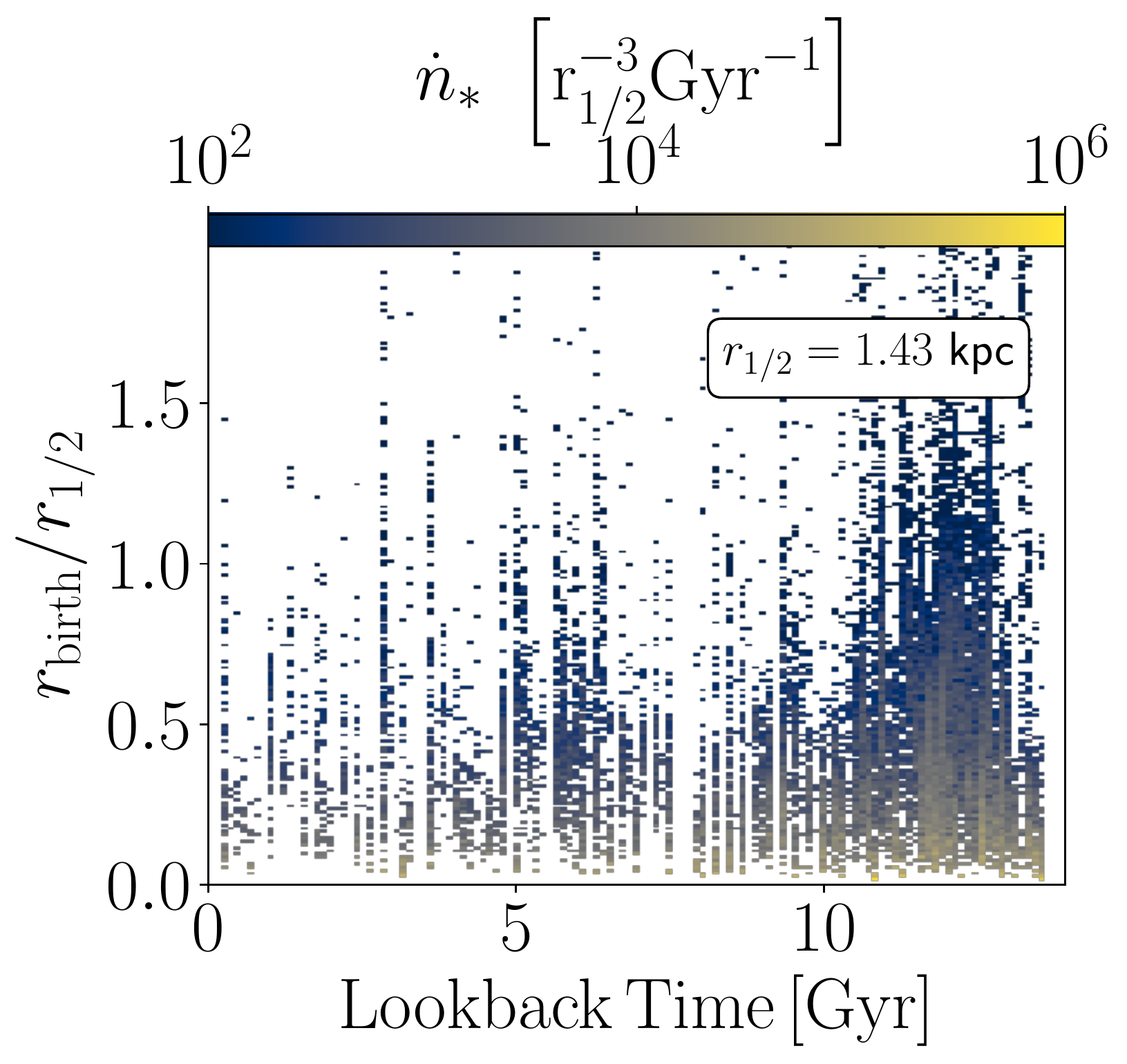}
    \includegraphics[width=.3\textwidth]{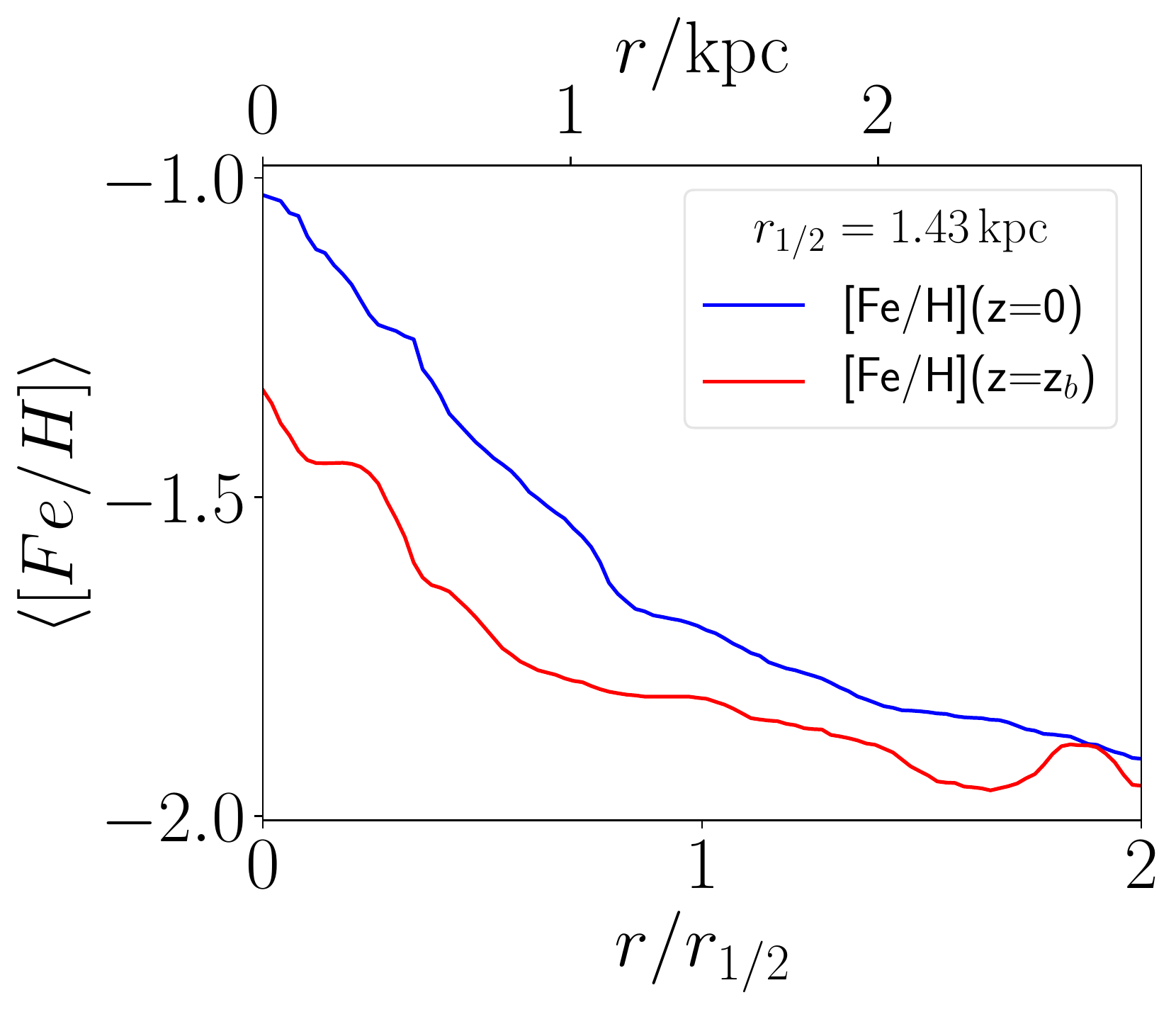}
    \includegraphics[width=.3\textwidth]{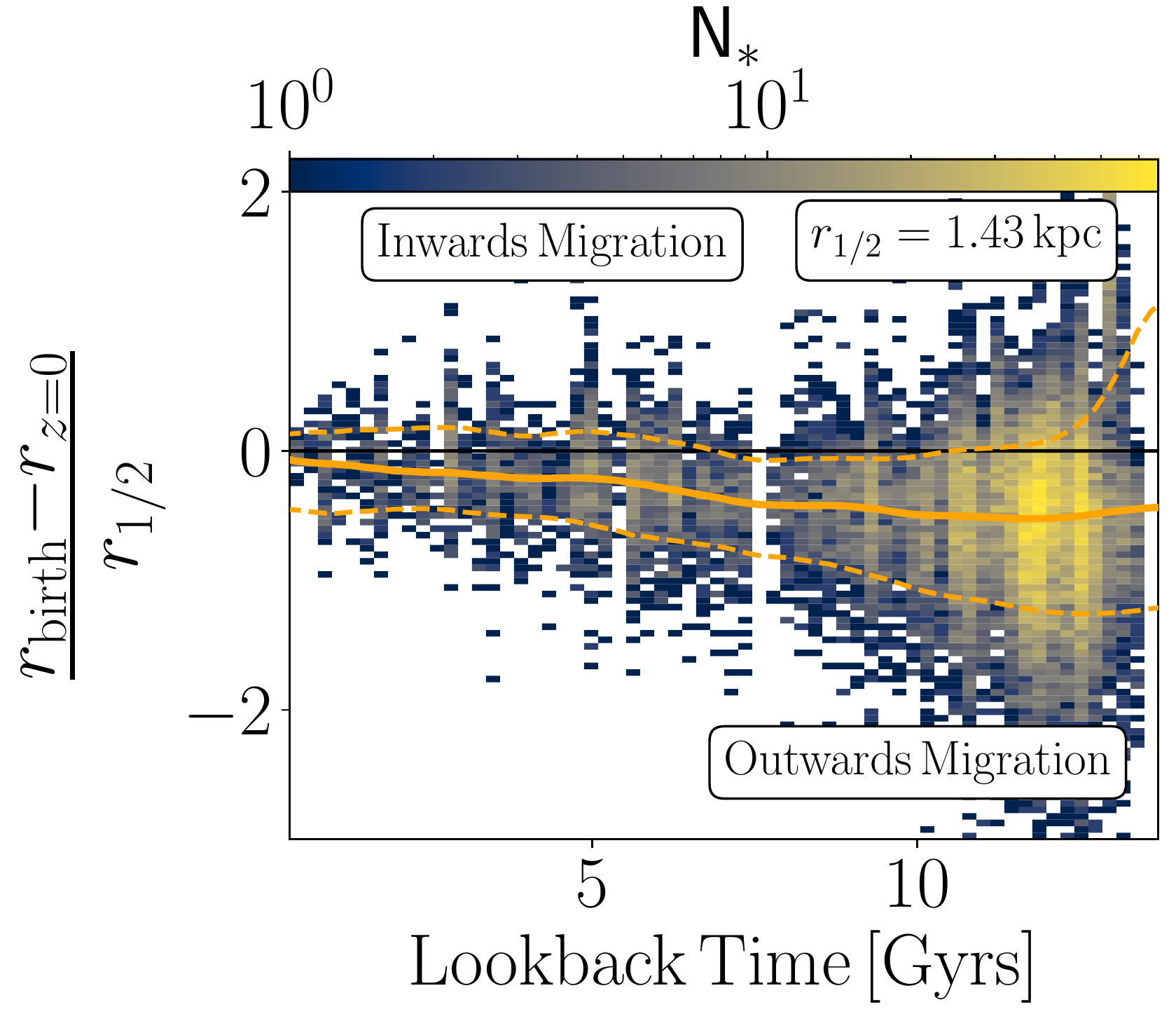}
    \caption{Evolution of the metallicity profile due to stellar redistribution in simulated UDGs. 
    Top panels refer to a rotation supported UDG, while bottom panels refer to a dispersion supported UDG. 
    Left panels show the evolution of the star formation rate  profile (i.e. the number of star particles formed at a given position and time in a unit volume and unit time). 
    In the middle panels we show the $3$D median metallicity  profile as a function of the radial position of the stars at $z=0$ (blue) and the radial positions of the stars when they were born (red). 
    The profiles have been built using a ruining median (weighted by a cubic spline kernel of width $0.2{\rm r}_{1/2}$).
    If the stellar component does not move from its birth position, the $z=0$ metallicity profile and the birth metallicity profile should match. The birth metallicity profile is strongly correlated to the radial distribution of the star formation (central panel). Furthermore, stellar redistribution plays a fundamental role in setting the $z=0$ metallicity profiles, in right panels we show a $2D$ histogram of the displacement of the stars, measured as the difference between birth radius and $z=0$ radius (normalized by ${\rm r}_{1/2}$), and the formation time. Orange solid lines are a ruining mean of the the displacement, while the dashed lines indicate the $1\sigma$ deviation.  These histograms  show how the stellar displacement happens smoothly with age, being the older star the ones that have move further from their birth positions. 
    }
    \label{fig:BirthProfile}
\end{figure*}
The absence of a linear correlation between the NIHAO UGDs stellar metallicity gradients and the median age, together with the moderate correlation with the half-mass radius,  suggests that  the relation between the feedback "puffing" and the metallicity gradients must be hidden by another process that dilutes it. 

Here we will explore correlations with other physical properties of the simulated UDGs, in order to disentangle what drives the metallicity gradient distribution in NIHAO UDGs. 

In Fig.~\ref{fig:Grad_CCoded} we show the metallicity gradient as a function of the age and half-mass radius (same as in Fig.~\ref{fig:SlopeAge_rel}, and for an aperture of $2{\rm R}_{1/2}$) color coded by the stellar mass (top),  the time since the last major merger (middle) and a parameter quantifying the kinematical support of each galaxy (bottom).
This parameter, $\kappa_{co}$, is a $3$D quantity that measures the contribution of the ordered circular motion compared to the total kinetic energy of the galaxy. The parameter is defined between $0$ and $1$, and larger values means a stronger rotation support.
More in detail, to estimate the rotation support of the galaxies we measured the fraction of kinetic energy invested in ordered rotation \citep{Correa2017}, which is defined as:

\begin{equation}\label{eq:kappa}
    \kappa_{\rm co} = \frac{K_{\rm rot}}{K} = \frac{1}{K}\sum_{i}\left[\frac{1}{2}m_{i}\left(\frac{l_{\parallel}^{i}}{R_{\bot}^{i}}\right)^2 
    \Theta\left(l_{\parallel}^{i}\right)\right],
\end{equation}

where $K = \frac{1}{2}\sum_i m_i v_i^2$ is the total kinetic energy, $l_{\parallel}^{i}$ the specific angular momentum parallel to the rotation axis, $m_i$  the mass of the $i$-th particle, $R_{\bot}^{i}$ the distance between the $i$-th particle and the rotation axis. These sums are aperture dependent so we used all the stellar particles inside $0.2R_{200}$. In order to take into account only co-rotating particles the summation for $K_{\rm rot}$ has been only done over particles with positive angular momentum \footnote{We make this explicit in Eq.~\eqref{eq:kappa} by including the step function $\Theta(x)$, which is $0$ if $x < 0$  and $1$ elsewhere.}.

Analyzing Fig.~\ref{fig:Grad_CCoded}, It seems that more massive UDGs tend to have flatter metallicity gradients than low mass UDGs (top panel of  Fig.~\ref{fig:Grad_CCoded}). Their larger mass implies  a larger ${\rm R}_{1/2}$  being thus partially responsible of the $\gamma_{\rm [Fe/H]} - {\rm R}_{1/2}$ correlation. Furthermore, massive systems are more likely to have undergone a recent merger\footnote{We have built  merger trees of NIHAO UDGs with the Merger-Tree feature of the Amiga Halo Finder \citep{AHF2004,AHF2009}, \url{http://popia.ft.uam.es/AHF/}} (middle panel of Fig.~\ref{fig:Grad_CCoded}). Here, we consider a merger any halo that crosses the  virial radius of the main progenitor with  a peak virial mass ratio larger than $10\%$, and we select as merger time, the snapshot in which such mass ratio is maximum (this generally happens just before the accreted progenitor falls into the virial radius of the main halo).
These  mergers tend to add energy to the pre-existing stellar component and, as a result,  moving it to the outskirts. If there is star formation after the mergers this will create a two component galaxy with a centrally concentrated young stellar population and an envelope of older stars, leading to steep metallicity gradients \citep[][]{BenitezLlambay2016, Cardona-Barrero2021}. However, if there is no enough time (or the SF rate is low) to develop such inner population of metal rich stars, the effect of the merger will be to effectively vanish pre-existing gradients.

Another mechanism that is known to affect the metallicity gradients is the amount of rotational support that both the stars and  gas have \citep[][]{Schroyen2011}. Rotation supported systems are more likely to have flatter or even positive gradients, as indicated  in the bottom panels of Fig.~\ref{fig:Grad_CCoded}, and this is likely due to the centrifugal barrier that forces them to have more spatially  extended star formation.
\cite{CardonaBarrero2020} showed that rotation supported UDGs and dispersion supported ones evolved in different ways depending on the alignment of the accreted baryons, leading to a disk-like and more triaxial populations respectively: rotation supported UDGs will therefore have larger sizes, as shown in the bottom right panel of Fig.~\ref{fig:Grad_CCoded}. Moreover, \cite{CardonaBarrero2020} found that the SF history of the rotation and dispersion supported populations are similar, meaning that despite having a different evolution, the age distribution of these two populations should be similar, as can   be appreciated in the bottom-left panel of Fig.~\ref{fig:Grad_CCoded}.  
Overall, Fig.~\ref{fig:Grad_CCoded} suggests that  UDGs in which the rotation support is higher populate a different region in the metallicity slope vs age diagram compared to the dispersion supported UDGs.  This is likely the main difference between our results and those of \cite{Mercado2021}, having them mostly dispersion supported galaxies.

In  Appendix~\ref{apx:mixture} we explore this by using a Bayesian approach, in order to check whether the dispersion and rotation supported UDGs indeed populate statistically well differentiated regions on the metallicity gradient versus $t_{50}$ diagram. We assume that there exist two outcomes of UDGs, one in which the metallicity gradient is linearly related to the median age of the stars and a second one (which we treat as contamination or outliers from the main relation) in which the metallicity gradient is set by the stellar kinematics. This approach  indicates that there is a tight relation between the metallicity gradient and the median age of the stars, and that such relation is mainly driven by the old ($t_{50}>7~{\rm Gyr}$) and dispersion supported ($\kappa_{\rm co} < 0.4$) population of UDGs. On the other hand, old UDGs with flat metallicity gradients are strongly identified as outliers (with posterior probabilities larger than $>95\%$), strongly favoring the conclusion that the metallicity gradient of these systems has been flattened due to the kinematics. Moreover, young ($t_{50} < 7~{\rm Gyr}$) and dispersion supported UDGs seem not to be well constrained in the model, they deviate from the most likely metallicity gradient vs median age relation towards flatter profiles and do not have strong probability of belonging to any of the two populations. Interestingly, these unconstrained galaxies are more likely to have undergone a recent merger suggesting that the role of the late mergers is to flatten the metallicity gradient by mixing the stellar populations, and the subsequent SF have not enough time to develop a gradient as steep as expected. 

\subsubsection{Evolution of  metallicity gradients}

In the previous section we have found rotation and dispersion supported UDGs to populate well differentiated regions in the $\gamma_{\rm [Fe/H]} - t_{50}$ diagram. In this section we will explore the origin of this dichotomy. 

As we have previously mentioned, the spatial segregation of stars with different metallicities is needed in order to build a  gradient. This can be achieved by different processes: the evolution of the locus of the star formation region and the spatial redistribution of different stellar populations due to SNae feedback \citep{DiCintio2017}. We explore these processes in Fig.~\ref{fig:BirthProfile}, taking as example two systems that evolve in complete isolation with no major mergers in the last $10$~Gyrs (i.e. without any merger with mass ratio larger than $1:10$). The top panels show a rotation supported UDG with flat metallicity gradient, while the bottom panels correspond to a dispersion supported UDG with steep metallicity gradient.

As we  anticipated in the previous section, a strong rotation  forces the gas to be distributed in a disk-like configuration, making the new stars  forming across the disk and further away from the center, as shown in the  top left panel of Fig.~\ref{fig:BirthProfile}, which represents the rate at which the stars form at any given time and distance from the center of the galaxy. This particular shape of the star formation history profile (known as "inside-out" evolution), is translated into an intrinsic positive metallicity gradient as old and metal poor stars remain in the center while new stars with higher iron abundances will be formed at larger distances. This is shown in the top central panel of Fig.~\ref{fig:BirthProfile}, in which the red and blue lines show the median $3$D iron abundance profile, as a function  of the radius at which the stars have born and of the radius at $z=0$, respectively. If the stellar component does not move from its birth position, the two  profiles should match: studying them therefore allows to characterize  the effect of stellar displacement due to SNae feedback.
In this case, we can observed that the $z=0$ profile is  flatter than the initial one. 

On the other hand, in the dispersion supported UDG (bottom panels), the gas is supported by pressure. In this galaxy the SF in the outskirts shuts off, in such a way that the new stars will form mostly in the center of the galaxy.  This process can be clearly seen in the bottom left panel of Fig.~\ref{fig:BirthProfile}, where we can see how the star formation is mostly confined in the inner half mass radius of the galaxy.This evolution of the SF region leads to a central region populated by a mixture of stars of different  ages and metallicities with a outer envelope of old and metal poor stars: this segregation naturally forms steep negative metallicity gradients as the one shown as a red line in the bottom central panel of Fig.~\ref{fig:BirthProfile}.

We explicitly show the displacement of the stars in the right panels of Fig~\ref{fig:BirthProfile}.  SNae feedback driven migration is a cumulative process, such that the older stars will suffer from more feedback cycles than the younger ones, moving to further distances.  This way, old stars, are effectively removed from the central regions and deposited at larger radii, increasing the central metallicity of the galaxy (see central panels of Fig.~\ref{fig:BirthProfile}). In rotation supported UDGs, this displacement of the old stars towards the surroundings of the galaxy will diminish the median metallicity of the outer parts of the disc, so that on average the $z=0$ metallicity profile will be effectively flattened in this case
(compare the red with blue lines of the top central panel of Fig.~\ref{fig:BirthProfile}).
On the other hand, in dispersion supported UDGs, the stars that are deposited into the outskirts by the stellar feedback, will encounter even older and less metallic stars. The average effect of this displacement is to increase the metallicity on the outskirts (compare the red with blue lines of the bottom central panel of Fig.~\ref{fig:BirthProfile}).

As a summary, the shape of the metallicity profile is initially settled by the way in which the SF region evolves, but it is strongly modified by subsequent feedback driven stellar displacement. 
The effect of stellar displacement due to feedback is to flatten the pre-existing positive metallicity gradients in rotation supported galaxies and to make steeper the negative ones in dispersion supported UDGs.

\section{Conclusions}
\label{sec:conclusions}

The large sizes  of  UDGs can be explained by SN feedback "puffing", as NIHAO simulations suggest \citep{DiCintio2017}. Indeed, SNae driven outflows generate strong perturbations in the gravitational potential of these galaxies, modifying the orbits of both stars and dark matter particles, and effectively creating cored, shallow distributions at their center \citep[e.g.][]{Governato2010, Duton2016}. 

In this contribution we explore the characteristics of stellar metallicity profiles in UDGs from the NIHAO simulations \citep{Wang2015} and compare them with a set of observed LG dwarfs \citep{Taibi} as well as of the only so-far observed metallicity profile in a UDG, i.e. the one of DF44 \citep{Villaume2022}.
A zero-th order approach is to check the average metallicity and age properties of simulated UDGs. NIHAO UDGs seem to have slightly smaller metallicities than LG dwarfs and UDGs found in clusters (Fig.~\ref{fig:Mass-Met_rel}, left panel), while  the few examples of field or nearly isolated UDGs, with reported spectroscopic metallicities, suggest that field UDGs are less metal rich, in agreement with NIHAO UDGs metallicities.  Moreover, both simulations and observations of LG dwarfs and several UDGs span a wide range of stellar ages (Fig.~\ref{fig:Mass-Met_rel}, right panel).

In order to explore possible gradients in metallicities, we first  built $100$ radial metallicity profiles per each galaxy, using  random orientations: we  found that the distribution of metals in simulated UDGs are best described by elliptical (instead of circular) apertures, and that normalizing the radial coordinate to the projected half mass radii of the profiles minimizes the variability due to projection effects (Fig.~\ref{fig:MetProfilesExm}). 

NIHAO UDGs show a large variety of metallicity profiles with slopes going from flat to negative (Fig.~\ref{fig:MetProfilesStacked}),  matching nicely the variability found in  LG dwarfs from \citet{Taibi}. However, only two simulated UDGs  seem to have a  shallow  flat-to-positive metallicity profile in their inner region, alike what observed  for DF$44$, as the reported in \cite{Villaume2022}. These two systems required a very recent SF triggered by late inflows of low metallicity gas, being likely not the case for DF$44$, since no recent star formation has been found for this galaxy. We note, however, that the data regarding DF44 show large error bars and are limited to a very small radial region well within the half-light radius of the galaxy. More extended radial profiles and a larger sample of observed metallicity in  UDGs  would be desirable in order to properly compare simulations and  observations. Indeed, the exact value of the average metallicity gradient is strongly dependent on the apertures used to measure it, as we show in top and bottom panels of Fig.~\ref{fig:SlopeAge_rel}.

Following \cite{Mercado2021}, we  explored the distribution of  metallicity slopes ($\gamma_{\rm [Fe/H]}$) vs ages ($t_{50}$) for our simulated UDGs.  When considering the full sample of simulated NIHAO UDGs we do not observe the relation between $\gamma_{\rm [Fe/H]}$ and $t_{50}$ found in \cite{Mercado2021}: this is likely due to the lack of rotation supported galaxies in their sample. 
However, when splitting the population of UDGs by their kinematics, we do recover an anti-correlation between  age and metallicity slope for the dispersion supported sample of galaxies, while rotation supported systems tend to have flat metallicity profiles despite being relatively old ($t_{50}>7~{\rm Gyr}$) galaxies (bottom panels of Fig.~\ref{fig:Grad_CCoded}). This dichotomy highlights the importance of the stellar kinematics in the properties of these systems. Moreover, our results are in agreement with the metallicity gradients found in the LG dwarfs as can be seen in Fig.~\ref{fig:SlopeAge_rel} \citep[][]{Taibi}.

Finally, we  found that the $z=0$ metallicity gradient distribution is mainly set by the interplay between the evolution of the  locus of the star formation region and the displacement of stars due to SNae feedback (Fig. \ref{fig:BirthProfile}). UDGs in which the SF region move to the outskirts (i.e. rotation supported systems) develop positive metallicity gradients, which get flattened by the subsequent stellar displacement, while  UDGs in which the SF region gradually shrinks towards the center with time (i.e. dispersion supported galaxies)  tend to develop intrinsic negative metallicity gradients,  which are steepened by stellar migration due to SNae feedback (see Fig.~\ref{fig:BirthProfile}).

These different evolving paths can be easily differentiated and constrained by the $z=0$ stellar kinematics and should serve as a test of our model, when more data will be available.

\section*{Acknowledgments}
S.C.B. acknowledges support from the Spanish Ministry of Economy and Competitiveness (MINECO) under the grant SEV-2015-0548-18-3 and the Spanish Ministry of Science and Innovation (MICIU/FEDER) through research grant PGC2018-094975-C22. A.D.C. is supported by a Junior Leader fellowship from `La Caixa' Foundation (ID 100010434), fellowship code  LCF/BQ/PR20/11770010.
G.B., S.C.B. and S.T. acknowledge support from the Agencia Estatal de Investigación del Ministerio de Ciencia en Innovación (AEI-MICIN) and the European Regional Development Fund (ERDF) under grant number PID2020-118778GB-I00/10.13039/501100011033 G.B. acknowledges the AEI under grant number CEX2019-000920-S.
S.T. acknowledges funding of a Leibniz-Junior Research Group (PI: M.~Pawlowski; project number J94/2020) via the Leibniz Competition.
This research was carried out on the High Performance Computing resources at New York University Abu Dhabi (UAE). 

Data analysis was performed using Python\footnote{\url{https://www.python.org}} programming language. The following Python modules were used for the analysis: pynbody \citep{pynbody}; pandas \citep{pandas}; numpy \citep{numpy}; scipy \citep{SciPy}; matplotlib \citep{matplotlib}; corner \citep{corner}; chainconsumer \citep{chainconsumer}; numba \citep{numba}; h5py \citep{h5py} and emcee \citep{emcee}.

\section*{Data Availability}
The data used in this work are available upon reasonable request to the
corresponding author and to the NIHAO PIs.



\bibliographystyle{mnras}
\bibliography{biblio.bib} 



\appendix

\section{Fitting probability distributions to median ages of LG dwarfs}\label{apx:fits}

In observations of dwarf galaxies, $t_{50}$ is difficult to constrain: the errors on this quantity are typically large, and in general highly asymmetric. We use the cumulative SFHs from \cite{Weisz2014} to obtain the  $16$-th, $50$-th $84$-th percentiles of $t_{50}$ (hereafter, $\tau_{16th},\tau_{50th},\tau_{84th}$). We thus only know the probability of $t_{50}$ being smaller than some pre-defined values , i.e. $P(t_{50} < \tau_{xx-th}) = xx/100$. In order to properly handle these data we have assumed  that $t_{50}$ follows a beta distribution, whose Probability Density Function (PDF, $f_{X}$) and Cumulative Distribution Function (CDF, $F_{X}$) can be written as:
\begin{align}
    f_{X}\left(x| a, b\right) &= \frac{x^{a-1}(1 -x)^{b-1}}{B(a, b)} \\
    F_{X}(x| a, b) &= \frac{B(x; a, b)}{B(a, b)} 
\end{align}

with $B(a, b)$ and $B(x; a, b)$ being the beta and the incomplete beta function, respectively:
\begin{align}
    B(x; a, b) &= \int_{0}^{x}t^{a-1}(1-t)^{b-1}dt  \\
    B(a, b) &= B(1; a, b) = \frac{\Gamma(a)\Gamma(b)}{\Gamma(a+b)}.
\end{align}

Then the CDF can be fit to the percentiles provided by \cite{Weisz2014} by standard methods, finally obtaining the parameters $a, b$ of the best-fitting PDF. During the fitting procedure we have included the constrains that $a \ge 1$ and $b \ge 1$. This way we force $t_{50}$ to be smaller than the age of the Universe, and larger than $0$~Gyrs. Note that the beta distribution is only defined in the interval $[0, 1]$, so the data should be normalized before performing the fit.

Once the best-fitted PDF of each galaxy is obtained, we can randomly sample one value of $t_{50}$ and perform the linear fit to this re-sampled data-set. Repeating this procedure leads us to obtain not only the best fitted linear relation but also its confidence intervals. 

We have applied this re-sampling to both $t_{50}$, assuming a beta distribution, and to the metallicity gradient, assuming a Gaussian distribution with $\sigma$, the error of the linear fit to the metallicity profiles of \cite{Taibi}. 

\begin{figure}
    \centering
    \includegraphics[width=1\columnwidth]{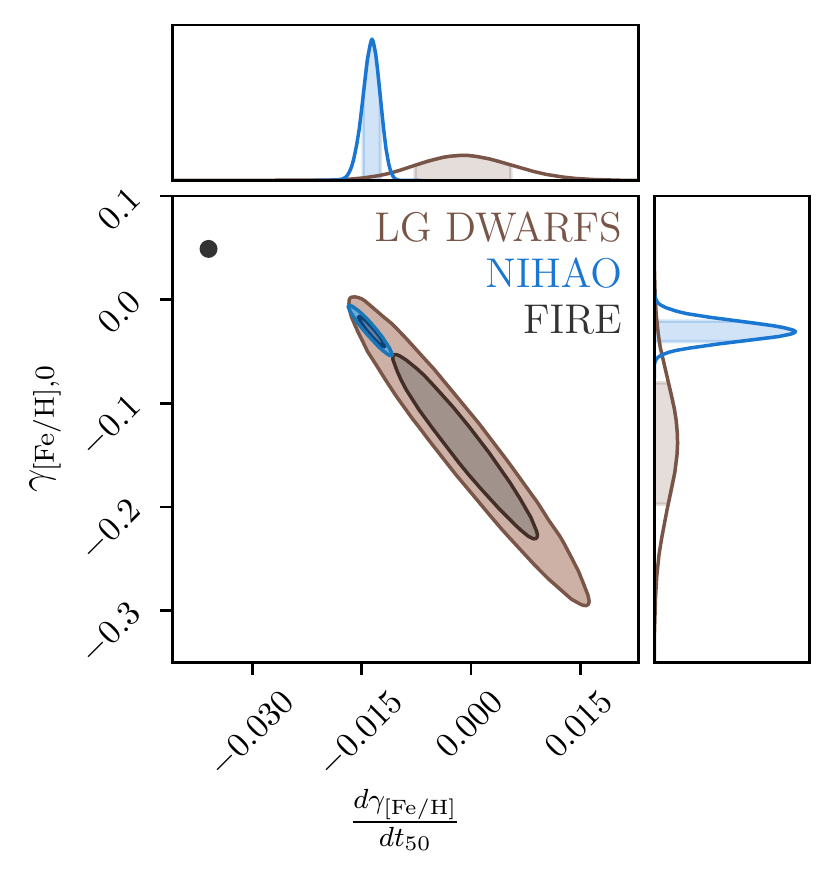}
    \includegraphics[width=1\columnwidth]{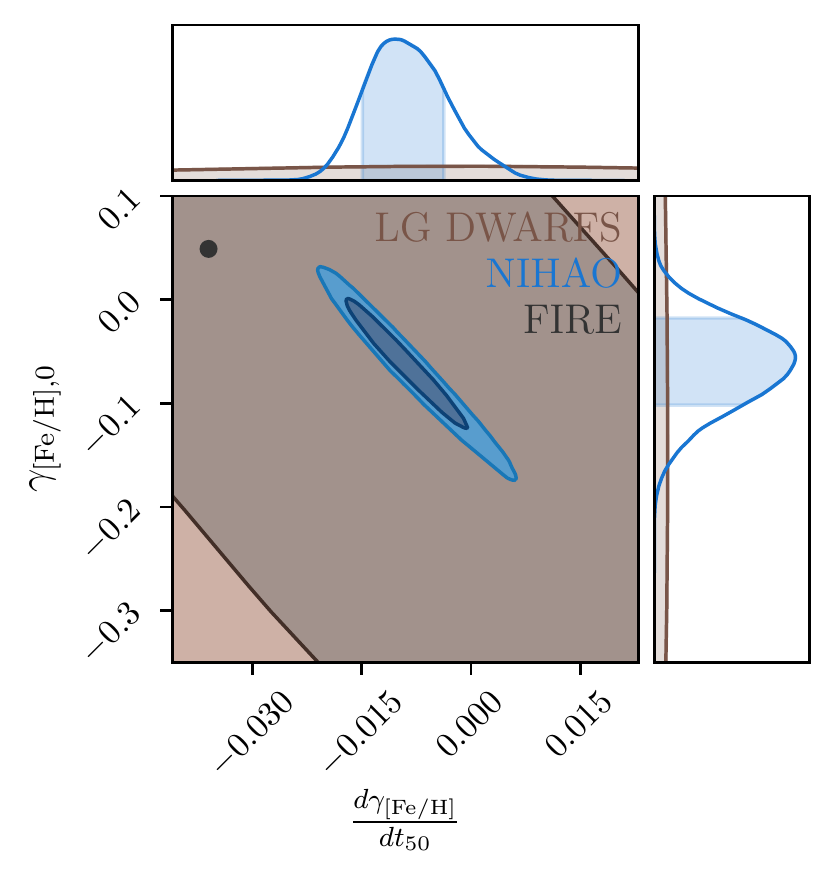}
    \caption{
    Best fit parameters for the linear relation between the metallicity gradient and median stellar age (Eq.~\ref{eq:lin}). The top and bottom panels show the parameters obtained when the metallicity slope is measured in an aperture of $2R_{1/2}$ and $0.6R_{1/2}$, respectively. The best-fitting  parameters for NIHAO UDGs and LG dwarfs are shown in blue and brown, respectively. The black circle shows the values  obtained by \protect\cite{Mercado2021} using FIRE simulations. }
    \label{fig:PosteriorFit}
\end{figure}

The resulting distribution of parameters for the fit between the metallicity slope and median age of the stars ($\alpha$ and $\beta$) can be found in Fig.~\ref{fig:PosteriorFit}.


\section{Mixture models}
\label{apx:mixture}
Here we perform a Bayesian analysis to derive an objective model to check whether the dispersion and rotation supported UDGs populate statistically well differentiated  regions on the metallicity gradient versus t$_{50}$ diagram.
This section closely follows the derivations of the likelihood for mixture models of \citet{Hogg2010, foreman_mackey_2014_15856}.

We assume the data to follow a linear relation between the metallicity slope $y_{i}$ (averaged over all projections) and the age ($x_{i}$). We assume the uncertainty to be the standard deviation of the metallicity slope over the $100$ random projections  ($\sigma_{i}$). We assume the data follows a Gaussian likelihood as follows:
\begin{align}
    P(y_{i}&| x_{i}, \sigma_{i}, m, b, S, q_i=1) = \nonumber \\
    &\frac{1}{\sqrt{2\pi(\sigma_{i}^2 + S^2)}}\exp\left\lbrace-\frac{(y_{i} - (m x_{i}+b))^2}{2(\sigma_{i}^{2} + S^{2})}\right\rbrace,
\end{align}
with $m$ and $b$ the slope and y-intercept of the linear relation and $S$ the intrinsic scatter of the relation. 

We will also assume that some of the galaxies follow an alternative linear relation between the metallicity slope and the kinematics support ($\kappa_i$). We will characterize this relation using the following Gaussian likelihood:
\begin{align}
    P(y_{i}&| \kappa_{i}, \sigma_{i}, MK, B, V, q_i=0) = \nonumber \\
    &\frac{1}{\sqrt{2\pi(\sigma_{i}^2 + V^2)}}\exp\left\lbrace-\frac{(y_{i} - \left(M \kappa_{i} + B\right))^2}{2(\sigma_{i}^{2} + V^{2})}\right\rbrace, 
\label{eq:likelihood}
\end{align}
being $M$, $B$ and $V$ the slope, zero point and intrinsic scatter of the relation between the metallicity gradient and the kinematics. 

We highlight the inclusion of the $q_i$ binary flag, which indicates if a given data point is drawn from the foreground model ($q_i=1$) or from the background one ($q_i=0$). The full likelihood is then:
\begin{equation}
    P(\lbrace y_{i}\rbrace| x_{i}, \sigma_{i}, m, b, S,  MK, B, V, \lbrace p_i\rbrace)= \prod_{i=1}^{N} P(y_{i}| x_{i}, \sigma_{i}, \theta, q_i)
\end{equation}
with $\theta$ a shortcut for appropriate  vector parameters.

These are thus $6+N$ parameters.  We can reduce the number of dimensions by marginalizing over the ${q_i}$. For that we shall introduce a prior over the $q_i$:
\begin{equation}
    P(p_i) = \left\lbrace
    \begin{array}{cc}
        Q & {\rm if}\, q_i = 1 \\
        1-Q & {\rm if}\, q_i = 0 
    \end{array}
    \right.
\end{equation}
Being in this case $Q$ the mean prior probability of a given point $i$ to be drawn from the foreground model. Multiplying the likelihood \eqref{eq:likelihood} by the previous prior, it can be shown that the marginalized likelihood can be written as:

\begin{equation}
\label{eq:post}
    \begin{aligned}
        P(\lbrace y_{k}\rbrace|& x_{k}, \sigma_{k}, m, b, S, B, V, Q) = \\
        \prod_{k=1}^{N}&\left[ 
        Q P(\lbrace y_{k}\rbrace| x_{k}, \sigma_{k}, m, b, S, p_k=1) + \right. \\
        &\left.(1-Q)P(\lbrace y_{k}\rbrace| x_{k}, \sigma_{k}, B, V, p_k=0)
        \right]
    \end{aligned}
\end{equation}

We have been finally left with a likelihood of only $7$ parameters.

We compute the posterior probability by including priors over our parameters. We used uniform priors for $b,\,B$ in the intervals: $(-0.5, 0.5),\,(-0.5, 0.2)$ respectively. For $V$ and $S$ we used log-uniform priors in the interval $-15 < \ln V < 0$ and $-9.2 < \ln S < 0$ . For $Q$ we used a flat prior in the interval $(0, 1)$.
Finally for the slopes of the linear relations we use the priors  $P(m) \propto (1 + m^2)^{-3/2}$ and $P(M) \propto (1 + M^2)^{-3/2}$ .

\begin{figure}
    \includegraphics[width=0.5\textwidth]{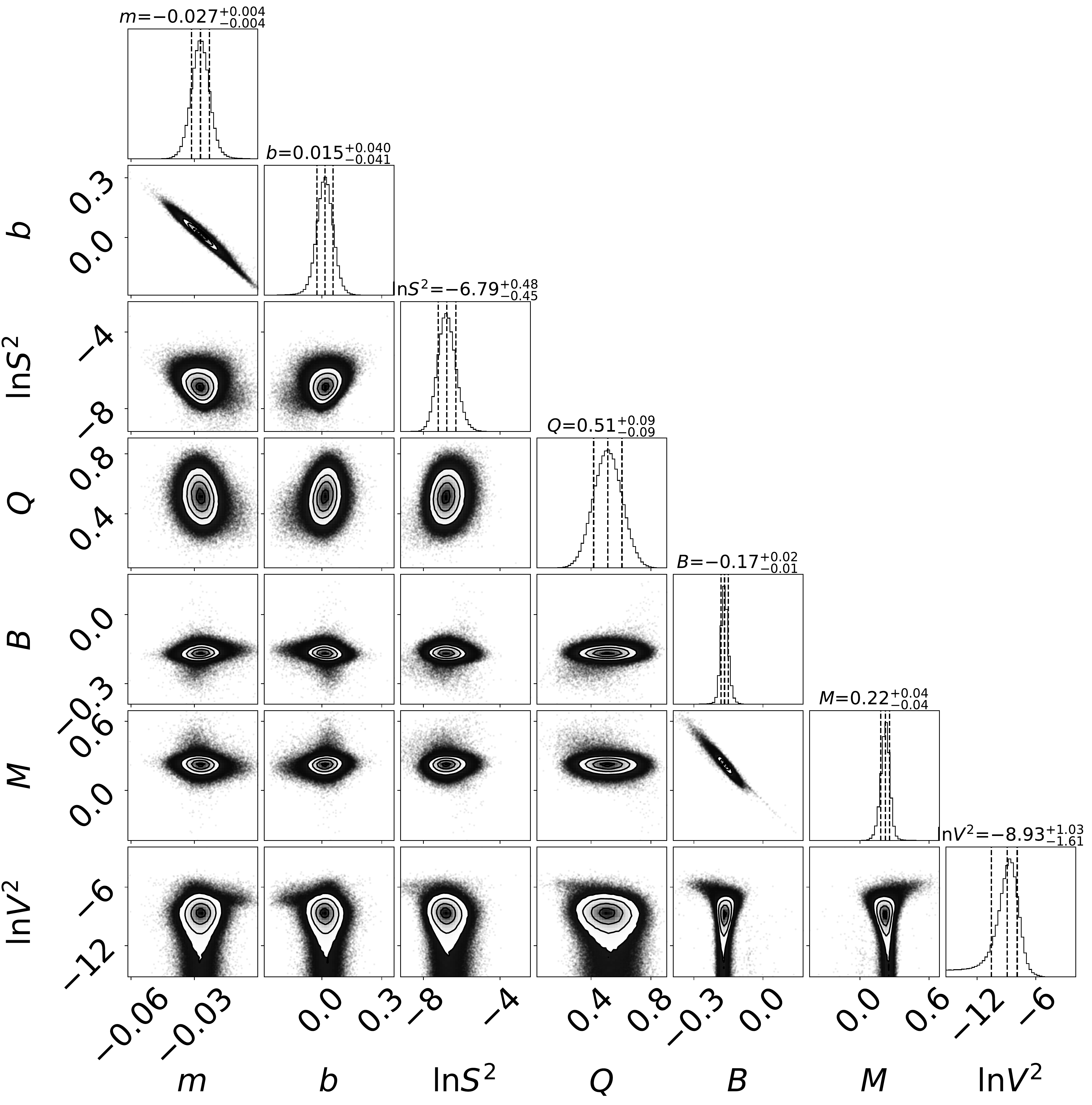}
    \caption{
    Samples obtained from the posterior probability distribution (Eq. \ref{eq:post}). Diagonal plot show the $1$D marginalized probability  of each parameter, while of diagonal panels show the marginalized $2$D posterior distributions for all the parameters. The titles indicated the median and the $16th$ and $84th$ percentiles over the marginalized distribution.}
    \label{fig:corner_plot}
\end{figure}

For sampling the posterior we have used $64$ walkers $1M$ iterations. We show the samples of the posterior distribution in Fig.~ \ref{fig:corner_plot}. 

\begin{figure}
    \centering
    \includegraphics[width=0.5\textwidth]{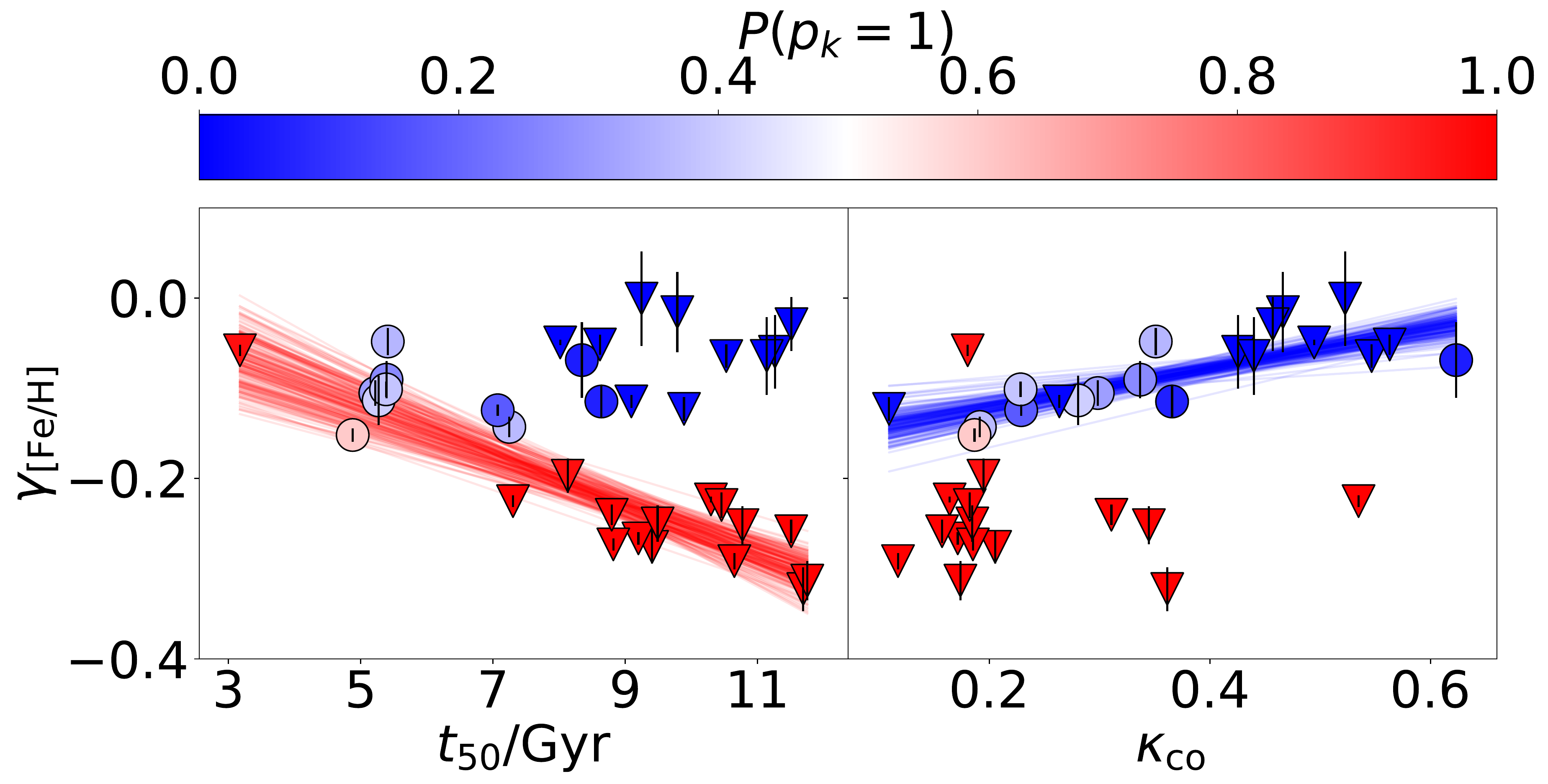}
    \caption{Metallicity slope vs Age(rotation support) relation on the left(right) color coded by the posterior probability of a point belonging to the foreground model (i.e. $p_k=1$). UDGs with posterior probabilities of belonging to the foreground(background) model larger than $0.95(0.05)$ are marked with inverted triangles.} 
    \label{fig:data_space}
\end{figure}

The projections over data space are shown in \ref{fig:data_space}. We have color coded the data points by their posterior probability of being drawn from the foreground model $p(q_{k}=1|y, \theta)$. We have used triangular symbols  on those data-points that show a posterior probability of being drawn from the foreground models larger than $0.95$ and smaller than $0.05$. 

\section{Tables}

\begin{table}
    \centering
     \caption{Best fit parameters for the metallicity slope vs median age relation. In this table we show the relation obtained using an aperture of $2$R$_{1/2}$. Each row shows the parameters obtained using different data-sets or methods. From top to bottom: relation obtained by \protect\cite{Mercado2021} using FIRE-2; relation obtained using LG-dwarfs from \protect\cite{Taibi}; NIHAO UDGs and NIHAO UDGs assuming an extra dependence with the stellar kinematics (see \ref{apx:mixture} for details)}
    \label{tab:model_params}
    \begin{tabular}{ccc}
        \hline
                Data-Set & $\frac{d\gamma_{\rm[Fe/H]}}{dt_{50}}$ & $\gamma_{\left[{\rm Fe/H}\right], 0}$ \\ 
                \hline \hline
                FIRE-$2$ & $\left(-36\pm5 \right) \times 10^{-3}$ & $0.049\pm0.035$  \\
                \hline
                Observed & $\left( -0.9^{+6.7}_{-6.4} \right) \times 10^{-3}$ & $-0.14^{+0.06}_{-0.06}$ \\ 
                NIHAO & $\left( -13.7^{+1.1}_{-1.2} \right) \times 10^{-3}$ & $-0.030^{+0.010}_{-0.010}$ \\ 
                \midrule
                NIHAO & $\left(-27.1^{+4.2}_{-4.3}\right)\times 10^{-3}$ & $\left.0.0152^{+0.0405}_{-0.0406}\right.$\\
                \bottomrule
    \end{tabular}

    \caption{Best fit parameters for the metallicity slope vs median age relation. In this table we show the relation obtained using an aperture of $0.6$R$_{1/2}$. Each row shows the parameters obtained using different data-sets, from top to bottom: LG-dwarfs from \protect\cite{Taibi} and NIHAO UDGs respectively.}
    \label{tab:model_params2}
    \begin{tabular}{ccc}
        \hline
                Data-Set & $\frac{d\gamma_{\rm[Fe/H]}}{dt_{50}}$ & $\gamma_{\left[{\rm Fe/H}\right], 0}$ \\ 
                \hline
                Observed & $\left( -3^{+62}_{-63} \right) \times 10^{-3}$ & $-0.15^{+0.46}_{-0.46}$ \\ 
                NIHAO & $\left( -9^{+6}_{-5} \right) \times 10^{-3}$ & $-0.062^{+0.039}_{-0.044}$ \\ 
                \hline
    \end{tabular}
\end{table}

\begin{table*}
 \centering
 \caption{From left to right: Galaxy ID; stellar mass; median iron abundance, median age of the stars, invested kinetic energy in ordered co-rotation \citep{Correa2017}; semi-major axis half-mass radius; semi-major axis half-light radius in the $r$-band; mean effective surface brightness in the $r$-band; metallicity slope computed in an aperture of $2{\rm R}_{1/2}$. 
 The values provided for half-mass radii, half-light radii, mean effective surface brightness and metallicity slope indicate the median over the $100$ different projections, while the errors the indicate $16$th and $84$th percentiles. We provide for the iron abundance and median stellar age the $16$th and $84th$ percentiles of their respective distributions.}
 \begin{tabular}{lrrrrrrrc}
 \toprule 
 ID & $\log_{10}{\rm M}_{\ast}/{\rm M}_{\odot}$ & [Fe/H] & $t_{50}$ [Gyr] & $\kappa_{\rm co}$ & ${\rm R}_{1/2}$ & ${\rm R}_{\rm eff}^{r}$ & $\left<\mu_{\rm eff}^{r}\right>$ & $\gamma_{\rm [Fe/H]}\,\left[{\rm dex\cdot}{\rm R}_{1/2}^{-1}\right]$ \\ 
\midrule \\ [-1em]
 g1.05e11 & $8.761$ & $-1.24_{-0.47}^{+0.22}$ & $5.22_{-3.33}^{+5.44}$ & $0.298$ & $4.85_{-0.90}^{+0.58}$ & $4.17_{-0.85}^{+0.65}$ & $23.02_{-0.29}^{+0.41}$ & $-0.106_{-0.014}^{+0.017}$ \\ [.5em] 
 g1.08e11 & $8.934$ & $-1.02_{-0.68}^{+0.23}$ & $9.79_{-4.34}^{+2.12}$ & $0.466$ & $4.16_{-0.11}^{+0.12}$ & $4.32_{-0.18}^{+0.19}$ & $23.56_{-0.31}^{+0.40}$ & $-0.006_{-0.067}^{+0.037}$ \\ [.5em] 
 g1.09e10 & $6.822$ & $-1.98_{-0.76}^{+0.43}$ & $11.69_{-11.47}^{+0.83}$ & $0.361$ & $1.46_{-0.12}^{+0.08}$ & $1.00_{-0.06}^{+0.05}$ & $24.63_{-0.47}^{+0.28}$ & $-0.324_{-0.026}^{+0.029}$ \\ [.5em] 
 g1.44e10 & $6.826$ & $-1.75_{-0.70}^{+0.30}$ & $10.53_{-7.41}^{+1.29}$ & $0.547$ & $1.56_{-0.18}^{+0.10}$ & $1.45_{-0.16}^{+0.11}$ & $26.11_{-0.58}^{+0.33}$ & $-0.061_{-0.016}^{+0.008}$ \\ [.5em] 
 g1.52e11 & $8.954$ & $-1.22_{-0.64}^{+0.29}$ & $8.34_{-4.69}^{+3.69}$ & $0.623$ & $6.24_{-0.72}^{+0.55}$ & $6.37_{-0.96}^{+0.66}$ & $24.03_{-0.61}^{+0.49}$ & $-0.060_{-0.057}^{+0.034}$ \\ [.5em] 
 g1.57e11 & $9.071$ & $-1.14_{-0.38}^{+0.30}$ & $5.27_{-3.12}^{+3.71}$ & $0.281$ & $5.28_{-1.02}^{+0.47}$ & $5.11_{-1.04}^{+0.47}$ & $23.10_{-0.31}^{+0.34}$ & $-0.111_{-0.035}^{+0.031}$ \\ [.5em] 
 g1.59e11 & $8.840$ & $-1.26_{-0.50}^{+0.32}$ & $9.25_{-4.21}^{+2.03}$ & $0.523$ & $6.22_{-0.42}^{+0.20}$ & $6.77_{-0.56}^{+0.28}$ & $24.72_{-0.78}^{+0.33}$ & $0.026_{-0.091}^{+0.015}$ \\ [.5em] 
 g1.88e10 & $7.217$ & $-1.83_{-0.39}^{+0.43}$ & $11.26_{-4.06}^{+1.03}$ & $0.426$ & $1.68_{-0.11}^{+0.07}$ & $1.69_{-0.14}^{+0.07}$ & $25.76_{-0.57}^{+0.33}$ & $-0.045_{-0.058}^{+0.026}$ \\ [.5em] 
 g1.89e10 & $7.114$ & $-1.74_{-0.31}^{+0.12}$ & $3.17_{-2.43}^{+2.52}$ & $0.180$ & $1.58_{-0.33}^{+0.21}$ & $1.39_{-0.30}^{+0.17}$ & $24.23_{-0.40}^{+0.24}$ & $-0.058_{-0.007}^{+0.005}$ \\ [.5em] 
 g1.90e10 & $7.081$ & $-1.71_{-0.50}^{+0.30}$ & $7.25_{-3.88}^{+2.10}$ & $0.192$ & $1.53_{-0.32}^{+0.15}$ & $1.36_{-0.27}^{+0.14}$ & $25.12_{-0.32}^{+0.21}$ & $-0.142_{-0.014}^{+0.012}$ \\ [.5em] 
 g2.09e10 & $6.973$ & $-1.77_{-0.65}^{+0.38}$ & $10.77_{-4.55}^{+1.34}$ & $0.345$ & $1.94_{-0.31}^{+0.26}$ & $1.89_{-0.30}^{+0.25}$ & $26.25_{-0.31}^{+0.30}$ & $-0.257_{-0.008}^{+0.015}$ \\ [.5em] 
 g2.34e10 & $7.136$ & $-1.89_{-0.23}^{+0.56}$ & $9.41_{-4.44}^{+1.22}$ & $0.205$ & $1.76_{-0.30}^{+0.20}$ & $1.52_{-0.24}^{+0.17}$ & $25.33_{-0.24}^{+0.22}$ & $-0.281_{-0.011}^{+0.028}$ \\ [.5em] 
 g2.37e10 & $7.251$ & $-1.71_{-0.30}^{+0.24}$ & $8.62_{-4.86}^{+2.79}$ & $0.563$ & $3.45_{-0.45}^{+0.22}$ & $1.24_{-0.17}^{+0.07}$ & $24.57_{-0.28}^{+0.32}$ & $-0.052_{-0.010}^{+0.011}$ \\ [.5em] 
 g2.63e10 & $7.633$ & $-1.74_{-0.38}^{+0.70}$ & $11.75_{-4.22}^{+0.96}$ & $0.174$ & $1.30_{-0.20}^{+0.18}$ & $1.17_{-0.18}^{+0.17}$ & $24.06_{-0.20}^{+0.15}$ & $-0.314_{-0.023}^{+0.024}$ \\ [.5em] 
 g2.64e10 & $7.473$ & $-1.76_{-0.35}^{+0.19}$ & $8.02_{-2.63}^{+4.14}$ & $0.495$ & $2.46_{-0.22}^{+0.11}$ & $2.42_{-0.21}^{+0.10}$ & $25.98_{-0.43}^{+0.31}$ & $-0.049_{-0.004}^{+0.003}$ \\ [.5em] 
 g2.80e10 & $7.567$ & $-1.53_{-0.40}^{+0.22}$ & $9.20_{-5.50}^{+2.09}$ & $0.171$ & $2.20_{-0.50}^{+0.30}$ & $1.89_{-0.44}^{+0.24}$ & $24.66_{-0.32}^{+0.29}$ & $-0.267_{-0.006}^{+0.008}$ \\ [.5em] 
 g2.83e10 & $7.471$ & $-1.75_{-0.40}^{+0.40}$ & $9.89_{-2.80}^{+2.16}$ & $0.109$ & $1.77_{-0.04}^{+0.03}$ & $1.72_{-0.04}^{+0.03}$ & $25.36_{-0.33}^{+0.32}$ & $-0.122_{-0.013}^{+0.013}$ \\ [.5em] 
 g2.94e10 & $7.771$ & $-1.54_{-0.68}^{+0.37}$ & $10.30_{-3.83}^{+1.79}$ & $0.164$ & $2.11_{-0.42}^{+0.19}$ & $1.84_{-0.35}^{+0.16}$ & $24.33_{-0.28}^{+0.17}$ & $-0.223_{-0.004}^{+0.003}$ \\ [.5em] 
 g3.23e11 & $8.562$ & $-1.32_{-0.30}^{+0.19}$ & $5.39_{-3.16}^{+4.01}$ & $0.337$ & $4.77_{-0.68}^{+0.29}$ & $4.19_{-0.54}^{+0.39}$ & $23.71_{-0.34}^{+0.33}$ & $-0.096_{-0.016}^{+0.032}$ \\ [.5em] 
 g3.44e10 & $7.803$ & $-1.53_{-0.53}^{+0.17}$ & $7.07_{-4.33}^{+4.52}$ & $0.229$ & $2.60_{-0.35}^{+0.20}$ & $2.29_{-0.35}^{+0.17}$ & $24.13_{-0.27}^{+0.34}$ & $-0.123_{-0.008}^{+0.006}$ \\ [.5em] 
 g3.67e10 & $7.740$ & $-1.70_{-0.48}^{+0.48}$ & $11.51_{-2.77}^{+0.93}$ & $0.157$ & $1.90_{-0.25}^{+0.12}$ & $1.75_{-0.23}^{+0.11}$ & $24.48_{-0.19}^{+0.26}$ & $-0.259_{-0.014}^{+0.014}$ \\ [.5em] 
 g3.93e10 & $7.576$ & $-1.65_{-0.44}^{+0.18}$ & $9.09_{-4.74}^{+2.14}$ & $0.264$ & $3.13_{-0.30}^{+0.36}$ & $2.68_{-0.19}^{+0.32}$ & $25.51_{-0.48}^{+0.48}$ & $-0.116_{-0.005}^{+0.011}$ \\ [.5em] 
 g4.27e10 & $7.801$ & $-1.63_{-0.33}^{+0.20}$ & $5.38_{-3.37}^{+5.98}$ & $0.228$ & $2.94_{-0.50}^{+0.20}$ & $2.35_{-0.37}^{+0.17}$ & $24.12_{-0.27}^{+0.23}$ & $-0.104_{-0.004}^{+0.009}$ \\ [.5em] 
 g4.48e10 & $8.147$ & $-1.60_{-0.16}^{+0.29}$ & $4.88_{-2.86}^{+5.04}$ & $0.187$ & $3.87_{-0.76}^{+0.26}$ & $3.24_{-0.64}^{+0.23}$ & $24.00_{-0.32}^{+0.23}$ & $-0.154_{-0.005}^{+0.013}$ \\ [.5em] 
 g4.86e10 & $8.090$ & $-1.64_{-0.49}^{+0.35}$ & $11.51_{-6.56}^{+1.16}$ & $0.457$ & $2.22_{-0.12}^{+0.04}$ & $2.22_{-0.12}^{+0.05}$ & $24.36_{-0.55}^{+0.29}$ & $-0.017_{-0.048}^{+0.015}$ \\ [.5em] 
 g4.94e10 & $8.056$ & $-1.46_{-0.55}^{+0.35}$ & $8.82_{-5.12}^{+3.19}$ & $0.185$ & $2.39_{-0.44}^{+0.26}$ & $2.00_{-0.41}^{+0.21}$ & $23.61_{-0.33}^{+0.24}$ & $-0.273_{-0.007}^{+0.006}$ \\ [.5em] 
 g4.99e10 & $8.093$ & $-1.53_{-0.38}^{+0.41}$ & $8.80_{-4.51}^{+2.58}$ & $0.311$ & $2.72_{-0.36}^{+0.33}$ & $2.39_{-0.40}^{+0.36}$ & $24.07_{-0.38}^{+0.29}$ & $-0.241_{-0.013}^{+0.015}$ \\ [.5em] 
 g5.05e10 & $7.980$ & $-1.51_{-0.60}^{+0.35}$ & $10.65_{-3.63}^{+1.68}$ & $0.117$ & $1.95_{-0.31}^{+0.17}$ & $1.77_{-0.29}^{+0.17}$ & $23.84_{-0.26}^{+0.32}$ & $-0.292_{-0.010}^{+0.012}$ \\ [.5em] 
 g6.12e10 & $7.967$ & $-1.45_{-0.61}^{+0.32}$ & $9.49_{-6.66}^{+2.95}$ & $0.185$ & $2.32_{-0.21}^{+0.12}$ & $1.88_{-0.23}^{+0.15}$ & $23.82_{-0.16}^{+0.27}$ & $-0.252_{-0.019}^{+0.025}$ \\ [.5em] 
 g6.37e10 & $8.356$ & $-1.54_{-0.35}^{+0.33}$ & $7.30_{-4.84}^{+3.47}$ & $0.535$ & $4.36_{-0.48}^{+0.31}$ & $4.31_{-0.66}^{+0.43}$ & $24.13_{-0.50}^{+0.71}$ & $-0.224_{-0.010}^{+0.004}$ \\ [.5em] 
 g6.91e10 & $8.399$ & $-1.37_{-0.64}^{+0.41}$ & $10.46_{-3.05}^{+1.78}$ & $0.182$ & $2.52_{-0.36}^{+0.17}$ & $2.23_{-0.32}^{+0.16}$ & $23.39_{-0.38}^{+0.23}$ & $-0.225_{-0.022}^{+0.009}$ \\ [.5em] 
 g6.96e10 & $8.567$ & $-1.35_{-0.49}^{+0.36}$ & $8.13_{-4.42}^{+2.76}$ & $0.195$ & $3.75_{-1.07}^{+0.47}$ & $3.29_{-0.94}^{+0.34}$ & $23.16_{-0.37}^{+0.38}$ & $-0.202_{-0.017}^{+0.027}$ \\ [.5em] 
 g7.12e10 & $8.139$ & $-1.50_{-0.49}^{+0.35}$ & $11.14_{-3.40}^{+1.47}$ & $0.440$ & $2.64_{-0.08}^{+0.05}$ & $2.62_{-0.17}^{+0.08}$ & $24.33_{-0.78}^{+0.38}$ & $-0.042_{-0.085}^{+0.013}$ \\ [.5em] 
 g8.89e10 & $8.606$ & $-1.19_{-0.50}^{+0.18}$ & $8.64_{-5.58}^{+3.26}$ & $0.366$ & $3.01_{-0.28}^{+0.15}$ & $2.95_{-0.32}^{+0.19}$ & $23.10_{-0.40}^{+0.32}$ & $-0.114_{-0.022}^{+0.019}$ \\ [.5em] 
 g9.59e10 & $8.450$ & $-1.43_{-0.27}^{+0.17}$ & $5.41_{-3.26}^{+4.63}$ & $0.351$ & $4.71_{-0.37}^{+0.32}$ & $4.64_{-0.53}^{+0.31}$ & $24.08_{-0.32}^{+0.36}$ & $-0.049_{-0.014}^{+0.018}$ \\ [.5em] 
 \bottomrule 
 \end{tabular} 
 \end{table*} 


\bsp	
\label{lastpage}
\end{document}